\documentclass[a4paper,12pt]{article}
\usepackage[paperwidth=9.5in,paperheight=12.0in,
  left=1.0in,right=1.0in,top=1.0in,bottom=0.6in,includefoot,heightrounded]{geometry}
\usepackage[usenames,dvi
psnames]{pstricks}
\usepackage[utf8]{inputenc}
\pdfoutput=1
\usepackage{lmodern} 
\usepackage[T1]{fontenc}
\usepackage{multicol}
\usepackage{epsfig}
\usepackage[fleqn]{amsmath}
\usepackage{amsmath,amsfonts}
\usepackage{textcomp}
\usepackage{graphicx}
\usepackage{subcaption}
\usepackage{authblk}
\usepackage[font={small}]{caption}
\usepackage{floatrow}
\usepackage{epstopdf}
\usepackage[pdfencoding=auto, psdextra]{hyperref}
\usepackage{bookmark}
\usepackage[numbers,sort&compress]{natbib}
\hypersetup{pdfauthor={some author},pdftitle={eye-catching title}}
\usepackage {ifpdf}
\usepackage{multirow}
\usepackage{epsfig}
\usepackage{multicol}
\usepackage{array}
\newcolumntype{L}{>{\centering\arraybackslash}m{3cm}}
\usepackage{varioref}

\newcommand{\be}{\begin{equation}}
\newcommand{\ee}{\end{equation}}

\title{\textbf{Investigating Sterile Neutrino Flux in the Solar Neutrino Data}}
\author{\small{Ankush  
\footnote{ankush.bbau@gmail.com}}}
\author{\small{Rishu Verma 
\footnote{rishuvrm274@gmail.com}}}
\author{\small{Gazal 
Sharma\footnote{gazzal.sharma555@gmail.com}}}
\author{\small{B. C. Chauhan\footnote{chauhan@associates.iucaa.in}}}
\affil{\textit{Department of Physics \& Astronomical Science, }\\
\textit{ School of Physical \& Material Sciences,}\\
\textit{ Central University of Himachal Pradesh (CUHP),
Dharamshala, Kangra (HP),
India 176215}}
\date{}
\begin{document}

\maketitle

\begin{abstract}
There are compelling evidences for the existence of a fourth degree of freedom of neutrinos i.e. sterile neutrino. In the recent studies the role of sterile component of neutrinos has been found to be crucial, not only in particle physics, but also in astrophysics and cosmology. This has been proposed to be one of the potential candidates of dark matter. In this work we investigate the updated solar neutrino data available from all the relevant experiments including Borexino and KamLAND solar phase in a model independent way, and obtain bounds on the sterile neutrino component present in the solar neutrino flux. The mystery of the missing neutrinos is further deepening as subsequent experiments are coming up with their results. The energy spectrum of solar neutrinos, as predicted by Standard Solar Models (SSM), is seen by neutrino experiments at different parts as they are sensitive to various neutrino energy ranges. It is interesting to note that more than 98\% of the calculated standard model solar neutrino flux lies below 1 MeV. Therefore, the study of low energy neutrinos can give us better understanding and the possibility to know about the presence of antineutrino and sterile neutrino components in solar neutrino flux. As such, this work becomes interesting as we include the data from medium energy ($\sim$1 MeV) experiments i.e. Borexino and KamLAND solar phase. In our study we retrieve the bounds existing in literature, and rather provide more stringent limits on sterile neutrino ($\nu_{s}$) flux available in solar neutrino data.

\end{abstract}


\section{Introduction}
H. Bethe proposed that the thermonuclear reactions are responsible for energy generation inside the stars, and it was subsequently established too. Sun, being the closest star to us, glows due to these thermonuclear fusion reactions, which also produce a huge amount of neutrinos leaving the solar core without any hinderance. To look for these neutrinos, the era of the solar neutrino experiments began in the late 1960\textquotesingle s with the Homestake solar neutrino detector, which was headed by astrophysicists Raymond Davis, Jr. and his theoretician friend John N. Bahcall. This experiment was designed to understand the mechanism of fusion reactions taking place inside the sun, instead it discovered a significant shortfall of neutrino flux as compared to SSM. This is known  as the Solar Neutrino Problem (SNP) \cite{{Davis:1968cp},{Cleve:1998}}.  After a journey of about four decades a leading solution for the SNP was identified with the help of reactor experiment KamLAND \cite{Eguchi:2002dm} in Japan, which independently verified neutrino oscillations through Large Mixing Angle (LMA) \cite{LMA,LMA1,LMA2,LMA3,LMA4,LMA5,LMA6,LMA7}\bibliographystyle{unsrtnat}  
as the dominant solution using neutrinos from nearby reactors. Motivated by peculiar nature of neutrinos, a number of subsequent experiments were performed, which rather than solving the mystery of the missing neutrinos, deepened it further. It came to the light that more than 98\% of the solar neutrino flux lies below 1MeV. The rare $^{8}$B neutrino flux is the high energy tail of solar neutrinos for which statistically significant measurements have been made so far. As such, the explored part of the flux is just a tip of an iceberg, there could be more mysteries hidden deep inside the water. Along with LMA there might be some sub-dominant effect (s) also affecting the nature and dynamics of these so called ghost particles. Therefore, the study of low energy neutrino data can give us better understanding and the possibility to explore the presence of sterile neutrinos and other components in solar neutrino flux.

Neutrinos are special and have a unique status in the Standard Model (SM) in the sense that they are massless and interact solely through the weak interaction. They are left handed only, i.e. active and partner of a corresponding charged lepton in a weak isospin doublet. As such, there is no place for the right-handed neutrinos in the SM. Right-handed neutrinos, if they exist, would be weak isospin singlets, i.e. not sensitive to even weak interaction. They would communicate with the other SM member neutrinos (left-handed active neutrinos) through mixing only. That is the reason, the right-handed neutrinos are called as inactive i.e. sterile. 
There are mounting evidences for existence of additional degrees of freedom ( >3) of neutrinos, which indicate the existence of a new class of heavy and light sterile neutrinos, and possibility of their mixing with active ones \cite{sterile,sterile1,sterile2,sterile3,sterile4}\bibliographystyle{unsrtnat}.

The Stanford group has done analysis of the available solar neutrino data \cite{Sturrock:2004hx,Sturrock:2003kv,Caldwell:2003dw,Sturrock:2001qn}\bibliographystyle{unsrtnat} that provided increasing evidence that the neutrino flux from the sun is not constant but varies with well-known solar rotation periods. If such findings are confirmed ever in future, the need for an addition to the LMA solution will be obvious and will most likely rely on an interaction of the solar magnetic field with the neutrino  magnetic moment. Since an SFP (Spin Flavour Precession) conversion to active antineutrinos is unlikely, this interaction is expected to produce a significant and time varying flux of sterile neutrinos \cite{sterile}, \cite{Berezinsky:2002fa} and \cite{deHolanda:2002ma}. 
It has become evident that the mechanism of SFP to active antineutrinos in the sun is either absent or plays a subdominant role. 
In fact these active neutrinos would originate a sizeable $\bar\nu_e$ flux \cite{Eguchi:2003gg}, whose upper bound has become stricter and corresponds to 0.028\% of the $^8$B neutrino flux. 

In the light of latest data available from various solar neutrino experiments, like the Neutral Current Detectors (NCDs) phase of Sudbury Neutrino Observatory (SNO), SuperK-III, SuperK-IV, Borexino and KamLAND solar phase \citep{sno,ncd,sk,BorII,Bor,kl} for neutrinos of energy range $\sim$1 MeV, we derive, in a model independent way, bounds on the sterile neutrino component present in the solar neutrino flux. We update the limits on the sterile neutrinos $(\nu_s)$ flux and compare them with the previous results obtained using various SNO phase data and data from SuperKamiokande experiments. Various such analysis \cite{Barger:2001pf,Barger:2001zs,Barger:2002iv,Balantekin:2004hi,bmw,Band,kang,bcc,skumar,lalsingh1} have been done in the past, but our results include all the experimental data, in a way it is a global data analysis and we obtain more constrained limits on the sterile neutrino flux as compared to the bounds existing in literature. The degeneracy of SSM normalisation factor (f$_B$)  and active neutrinos (sin$^2 \alpha$) is updated and best fit values for various cases using $\chi^{2}$ analysis are obtained.

In section ({\bf\ref{sec:2}}) we present the experimental details and the data available from all the solar neutrino experiments. In section ({\bf\ref{sec:3}}) we discuss the entire theory of model independent analysis starting from the master equations. Section ({\bf\ref{sec:4}}) presents results and discussion of our work including the $\chi^{2}$-analysis for different combination cases of various experiments, and the conclusions are finally summarised in section ({\bf\ref{sec:6}}).

\section{Data from various Solar Neutrino Experiments} \label{sec:2}

Kamiokande, a large water Cerenkov detector, was primarily designed to do search for proton decay. It was diverted to take data of solar neutrinos in the second phase i.e. Kamiokande-II in 1985, and provided a directional information of solar neutrinos demonstrating directly for the first time that the sun is a source of neutrinos. Via elastic scattering (ES) experiment this also discovered the atmospheric neutrinos anomaly and observed neutrino signals from the supernova 1987A. 
Super-Kamiokande (SK) was designed after overhauling the Kamiokande to do search for proton decay, studied solar and atmospheric neutrinos, and watched the supernovae in the Milky Way Galaxy. 
SK is a cylindrical 50kt water Cerenkov detector which observes high energy solar neutrinos via ES of electrons. The SK experiment started taking data in April 1996 and continued the observation for five years till July 2001, the running period referred as SK-I. The SK detector was rebuilt after an accident met with the half of the original PMT density in the inner detector and resumed observation from October 2002, which is referred as the beginning of SK-II running period.
The SK-II continued the measurements for three years and finished in October 2005 for the reconstruction work to put the PMT density back to the SK-I level. The detector restarted observation in June 2006, which is referred as the SK-III period. 
The fourth phase of SK (SK-IV) started of in September of 2008, with new front-end electronics, QBEE (QTC Based Electronics with Ethernet)\cite{sk4,Yamada} for new data acquisition system.

The data obtained SNO consists also of high energy end part of $^{8}$B solar neutrinos. The SNO, based in Canada, detects neutrinos via three reaction channels 1) charged-current interactions (CC) on deuterons, in which only electron neutrinos participate; 2) neutrino-electron ES, which are dominated by contributions from electron neutrinos and 3) neutral-current (NC) disintegration of the deuteron by neutrinos, which has equal sensitivity to all active neutrino flavours. So far, the SNO has three stages/phases of running. The first phase was with pure D$_{2}$O (heavy water), ran from November 1999 to May 2001. In the second phase, to increase the neutron detection efficiency, 2000 kg of NaCl salt was added to the target material D$_{2}$O. This SNO-II (Salt phase) ran from June 2001 to October 2003. The third and final stage of SNO phase used NCDs, which was achieved by the removal of the salt and addition of 36 strings of $^{3}$He proportional counters  to provide an independent detection of neutrons. This SNO-III (NCDs)  phase ran from November 2004 to November 2006. As these three stages have very different systematic uncertainties for the detection of neutrons, the stages of the SNO (Phase -I, II, III) can be considered as three independent experiments in measuring the flux of $^{8}$B solar neutrino flux with the NC reaction. It is further added that in the second SNO phase of running a data set of 254.2 days with a 5.5 MeV energy threshold, and in the third phase of running for 385.2 days of data with a 5 MeV energy threshold are taken.

Kamioka Liquid Scintillating Anti-Neutrino Detector (KamLAND) is located in the Kamioka Mine near the city of Kamioka in the Gifu prefecture
of Japan. The reactor experiment KamLAND is exactly located in the old Kamiokande site within the mine. KamLAND started data collection in January 2002 and the results were reported using only 145 days of data. The detector consists of a 1000 metric tons (1 kilo-ton or 1 kt) of liquid scintillator as both the target and detection medium for low energy nuclear/particle physics processes like neutrino ES and inverse beta decay. The liquid scintillator molecules give off light when charged particles move through the detector. As such, the KamLAND is designed and instrumented to detect this light and reconstruct the physics processes that produce the light. In the solar phase, KamLAND is focussed to detect medium energy solar neutrinos ($\sim$1 MeV). We take the latest data for $^{8}$B solar neutrinos based
on a 123 kton-day exposure of KamLAND \cite {kl}. KamLAND also reported measurement of the neutrino-electron ES rate 862 keV $^7$Be solar neutrino based on a 165.4 kton-days exposure of KamLAND for 616 days between April 7, 2009 and June 21, 2011 \cite{bekam}.   

Borexino is an ultra-high radio purity detector located in the underground laboratory at Gran Sasso, Italy. It is a low threshold liquid scintillator detector, which detects solar neutrinos of medium energy range ($^{7}$Be, CNO, pep) via ES of electrons. This detector is the first experiment able to do a real-time analysis of low energy sector solar neutrinos. 
The modelling of the detector response has been steadily improved since the beginning of data taking in 2007, and invaluable information has been provided by an extensive calibration campaign in 2009. As a target, 300 ton of liquid scintillator is contained in a spherical nylon vessel. To provide a passive shielding, a non-scintillating buffer liquid is used outside. More than 2200 PMTs are mounted on the inner surface of a stainless steel sphere to register the scintillation light.  The water volume shield the detector against external gamma and neutron radiations, which acts as an active muon veto. 

Borexino reports the interaction rate of the 0.862 MeV $^{7}$Be solar neutrinos  for 192 live days data in the period from May 16, 2007 to April 12, 2008 (Borexino Phase-I). The first simultaneous measurement of the interaction rates of pp, $^{7}$Be, and pep solar neutrinos was taken in an extended energy range (0.19-2.93) MeV. This result pertains to the span of 1291.51 days, i.e. Borexino Phase-II data, which was collected between December 2011 and May 2016 after an extensive scintillator purification campaign. Borexino is the first experiment to succeed in suppressing all major backgrounds, above the 2.614 MeV photon ($\gamma$) from the decay of $^{208}$Tl, to a rate below that of electron scatterings from solar neutrinos. This allows to reduce the energy threshold for scattered electrons by $^8$B solar neutrinos to 3 MeV (Borexino (3 MeV)). Borexino also reported the $^8$B interaction rate with 5 MeV threshold (Borexino (5 MeV)).

The summary of SNO-I, II, III, SK-I, II, III, IV, SK-combined, KamLAND Solar Phase, Borexino (3 MeV) and Borexino (5 MeV) data are shown in the Table {\ref{Table:1}}. In Table {\ref{Table:medium}}, survival probabilities of $^7$Be solar neutrinos from KamLAND, Borexino Phase-I, Borexino Phase-II are given.  In Table {\ref{Table:2}} we present the data in terms of the corresponding rates with reference to the SSM flux \cite{SSM}.

\begin{table}[h]
\caption{Solar Neutrino Flux measured by various experiments in units of $10^6$ cm$^{-2}$s$^{-1}$.}
\label{Table:1}
\centering
	\begin{tabular}{|c|c|c|c|c|}
		\hline 
	\bf {Data} & \boldmath${\phi^{CC}}$ &  \boldmath${\phi^{NC}}$ & \boldmath$\phi^{ES}$ & \bf{Reference} \\ 
		\hline 
		SNO-I& 1.76 $\pm$ 0.11 & 5.09 $\pm$ 0.45 & 2.39 $\pm$ 0.27 & \cite{sno} \\ 
		\hline 
		SNO-II& 1.68 $\pm$ 0.10 &4.94 $\pm$ 0.41 &2.35 $\pm$ 0.27 & \cite{sno}\\
		\hline
		SNO-III& 1.67 $\pm$ 0.09 &5.54 $\pm$ 0.47 &1.77 $\pm$ 0.24 & \cite{sno}\\
		\hline
		SK-I& - & - &2.38 $\pm$ 0.08 & \cite{sk}\\
		\hline
		SK-II& - & - &2.41 $\pm$ 0.17 & \cite{sk}\\ 
		\hline 
		SK-III& - &-  &2.40 $\pm$ 0.07 & \cite{sk}\\ 
		\hline 
		SK-IV& - & - &2.308 $\pm$ 0.044 & \cite{sk}\\ 
		\hline 
		SK Combined& - & - &2.345 $\pm$ 0.04 & \cite{sk}\\
		\hline
		KamLAND&-  & - &2.77 $\pm$ 0.41 & \cite{kl}\\ 
		\hline 
		Borexino (3 MeV) &-  &-  &2.55 $\pm$ 0.20 & \cite{BorII}\\ 
		\hline
		Borexino (5 MeV) &-  &-  &2.70 $\pm$ 0.45 & \cite{Bor}\\ 
		\hline
	\end{tabular}
	\end{table}

\begin{table}[H]
		\centering
	\begin{tabular}{|c|c|c|}
		\hline 
	\bf {Experiment} & \bf{P}\boldmath{$_{ee}$} & \bf{Reference} \\ 
		\hline
		KamLAND  & 0.66 $\pm$ 0.15 & \cite{bekam}\\ 
		\hline 
		Borexino Phase-I & 0.51 $\pm$ 0.07 & \cite{Bor}\\ 
		\hline
		Borexino Phase-II & 0.53 $\pm$ 0.05 & \cite{bebor2}\\ 
		\hline
	\end{tabular}
\caption{Solar electronic neutrino survival probability measured by various experiments for $^7$Be neutrinos.}
\label{Table:medium}
\end{table}

\begin{table}[H]
\centering
\begin{tabular}{|c|c|c|c|}
	\hline 
	\bf{Experiment}  & \bf{R}\boldmath {$^{CC}$} & \bf{R}\boldmath{$^{NC}$} & \bf{R}\boldmath{$^{ES}$}  \\ 
	\hline 
	SNO I & 0.32 $\pm$ 0.04 & 0.93 $\pm$ 0.14 & 0.44 $\pm$ 0.07\\ 
	\hline 
	SNO II & 0.31 $\pm$ 0.04 & 0.90 $\pm$ 0.13 & 0.43 $\pm$ 0.07\\ 
	\hline 
	SNO III& 0.30 $\pm$ 0.04 & 1.01 $\pm$ 0.15 & 0.32 $\pm$ 0.05\\ 
	\hline 
	SK I&- & -  & 0.43 $\pm$ 0.05 \\
	\hline 
	SK II& - & - & 0.44 $\pm$ 0.06\\ 
	\hline 
	SK III& - &- & 0.44 $\pm$ 0.05 \\ 
	\hline 
	SK IV&- &   -& 0.42 $\pm$ 0.05\\ 
	\hline 
	SK Combined &  - & -& 0.42 $\pm$ 0.05\\
	\hline
	KamLAND&- &   - & 0.51 $\pm$ 0.09\\ 
	\hline 
	Borexino (3 MeV)& -  & - & 0.46 $\pm$ 0.06\\ 
	\hline 
	Borexino (5 MeV)& - &   -& 0.49 $\pm$ 0.10\\
	\hline
\end{tabular}
\caption{Different rates with errors for the SNO, SK, KamLAND and Borexino experiments.}
\label{Table:2}
\end{table}

\section{Model Independent Analysis Theory} \label{sec:3}
 In the master equations \cite{skumar,lalsingh1} of the model independent analysis given below we neglect electronic antineutrino component as evident from KamLAND results \cite{Eguchi:2003gg}
 \vspace{-1cm}
\begin{center}
\begin{eqnarray}
\phi^{CC}=\phi_{\nu_{e}} ,\label{eq:1}\\
\phi^{NC}=\phi_{\nu_{e}}+\phi_{\nu_{x}}+\bar{r}_{d} \phi_{\bar{\nu}_{x}},\label{eq:2}\\
\phi^{ES}=\phi_{\nu_{e}}+r\phi_{\nu_{x}}+\bar{r_{x}}\phi_{\bar{\nu}_{x}}.\label{eq:3}
\end{eqnarray}
\end{center}
From these equations we can see that the NC is sensitive equally for all neutrino components, the ES is more sensitive to electronic neutrino component than the non-electronic ones, and the CC is only due to the electronic component of the neutrinos.

 The quantities $r$, $\bar{r}_{x}$ used are the ratios of the NC neutrino and non-electronic antineutrino event rates to the NC + CC neutrino event rate, respectively. The remaining $\bar{r}_{d}$ is the ratio of the antineutrino deuteron fission to neutrino deuteron fission event rate. 
We calculate them using the following forms of expressions 
\begin{eqnarray}
r=\frac{\int dE_{\nu}\phi(E_{\nu})\int dE_{e}\int dE^{'}_{e}\frac{d\sigma_{NC}}
{dE_e}f(E^{'}_e,E_{e})}{\sigma_{NC}\rightarrow \sigma_{NC+CC}},\label{eq:4}
\end{eqnarray}
\begin{eqnarray}
\bar{r}_{x}=\frac{\int dE_{\nu}\phi(E_{\nu})\int dE_{e}\int dE^{'}_{e}\frac{d\bar\sigma_{NC}}
{dE_e}f(E^{'}_e,E_{e})}{\sigma_{NC}\rightarrow \bar\sigma_{NC+CC}},\label{eq:5}
\end{eqnarray}
\begin{eqnarray}
\bar r_{d}=\frac{\int dE_{\nu}\phi(E_{\nu})\bar\sigma_{NC}(E_{\nu})}
{\bar\sigma_{NC}\rightarrow \sigma_{NC}}.\label{eq:6}
\end{eqnarray}
Here the energy resolution function is $f(E^{'}_e,E_{e})$ and $\sigma$'s are the cross sections. 
 The subscript $x$ in $\nu_{x}$ and $\bar{\nu_{x}}$ stands for the non-electronic ($\mu/ \tau$) components. One can extract the expression for ES flux from the master equations \eqref{eq:1} - \eqref{eq:3} as
\begin{eqnarray}
\phi^{ES} = r\phi
^{NC}+(1-r)\phi^{CC}-(r \bar{r}_{d}-\bar{r}_{x})\phi_{\bar{\nu}_{x}}.\label{eq:7} 
\end{eqnarray}

If we neglect the antineutrino component, as there are stringent limits from KamLAND, the expression for ES flux is obtained as 
\begin{eqnarray}
\phi_{no \bar{\nu_{x}}}^{ES}=r\phi^{NC}+(1-r)\phi^{CC}. \label{eq:8}
\end{eqnarray}

 Since we are specifically interested in estimating the flux of sterile neutrinos, therefore we derive an expressions for the active neutrino ($\nu_{e}+\nu_{x}+ \bar{\nu}_{x}$), non-electronic neutrino component ($\nu_{x}$) and thereafter the sterile neutrino ($\nu_{sterile}$) fluxes can be estimated as
\begin{eqnarray}
\phi_{active}=\frac{[(r-\bar{r}_{x})\phi^{NC}+(1-\bar{r}_{d})((1-r)\phi^{CC}-\phi^{ES})]}{r\bar{r}_{d}-\bar{r}_{x}},\label{eq:9}\\
\phi_{\nu_{x}}^{NC}=\phi^{NC}-\phi^{CC}, \label{eq:10}\\  
\phi_{\nu_{x}}^{ES}=\frac{\phi^{ES}_{SK}-\phi^{CC}}{r}.\label{eq:11}
\end{eqnarray}

We can get the expression for the sterile neutrino flux if we subtract the active neutrino flux from the SSM predictions as
\begin{eqnarray}
\phi_{sterile}=\phi_{SSM}^{B}-\phi_{active}.\label{eq:12}
\end{eqnarray}




The active-sterile admixture can also be estimated as the mixing angle $\alpha$ between the active and sterile neutrinos, such that sin$^{2} \alpha$ will denote the fraction of all the active neutrinos. Therefore the component of the sterile neutrinos is proportional to cos$^2{\alpha}$. Excluding the electronic component the fraction of active neutrino flux, as measured by CC flux, present in the solar neutrino flux can be calculated by the following relation

\begin{eqnarray}
 \sin^{2}\alpha = \frac{\phi_{active}-\phi^{CC}}{\phi_{SSM}-\phi^{CC}}.\label{eq:13}
\end{eqnarray} 

The degeneracy between the sterile neutrino component and the parameter designated as f$_{B}$, the  normalisation to the SSM $^8$B neutrino flux, can be clearly seen if we rewrite the model independent flux equations in terms of the expressions for the CC, NC and ES rates \cite{bcc}
\begin{eqnarray}
R^{CC}=f_{B}P_{ee}, \label{eq:14}\\
R^{NC}=f_{B}P_{ee}+f_{B}(1-P_{ee})[\sin^{2}\alpha\sin^{2}\psi+\bar
{r}_{d}\sin^{2}\alpha\cos^{2}\psi],\label{eq:15} \\
R^{ES}=f_{B}P_{ee}+f_{B}(1-P_{ee})[r\sin^{2}\alpha\sin^{2}\psi+\bar{
r_{x}}\sin^{2}\alpha\cos^{2}\psi].\label{eq:16}
\end{eqnarray}
As evident from above equations, these are reduced rates.
Here we exploit the near energy independence in the range of interest and factorize the electron neutrino survival probability P$_{ee}$ out of these integrals as in equations \eqref{eq:14} - \eqref{eq:16}.  
The above equations clearly show that the conversion of electron neutrinos into the other flavours are directly proportional to 1-P$_{ee}$. Here sin$^{2}\psi$ gives the measure of antineutrino components. Since we assume negligible antineutrino components, it is safer to take sin$^{2}\psi = 1$.
Re-arranging the rate equations we get sin$^{2}\alpha$ and f$_{B}$ degeneracy equation as 
\begin{eqnarray}  
\sin^{2}\alpha=\frac{R^{ES}-R^{CC}}{(f_{B}-R^{CC}){r}}.\label{eq:19}
\end{eqnarray}

Using ES rate equation and the survival probability P$^{M}_{ee}$ for intermediate energy neutrinos, we obtain
\begin{eqnarray}
R^{ES}=P^{M}_{ee}+(1-P^{M}_{ee})[r\sin^{2}\alpha\sin^{2}\psi+\bar{r}_{x}\sin^{2}\alpha\cos^{2}\psi],\label{eq:20}
\end{eqnarray} 
where $r$ and $\bar{r}_{x}$ are cross-sectional ratios as defined above. From the above equation, rate for no antineutrino component (sin$^{2}\psi=1$) and a sterile admixture with the active neutrinos is given as
\begin{eqnarray}
R^{ES}_{BOR/KL}=P_{ee}^{M}+ (1-P_{ee}^{M})r\sin^{2}\alpha. \label{eq:21}
\end{eqnarray}

In both cases, the value of cross-sectional ratios $r$ will be different. 
As of today we know the corresponding ES rates ( R$^{ES}_{BOR}$ and  R$^{ES}_{KL}$) and the survival probabilities for medium energy neutrinos (P$^{M}_{ ee}$) from the experiments, therefore using these, we estimate the sterile neutrino flux component present in the observed data.

\section{ Results and Discussions} \label{sec:4}
\begin{figure}[H]
\centering
\includegraphics[scale=.50]{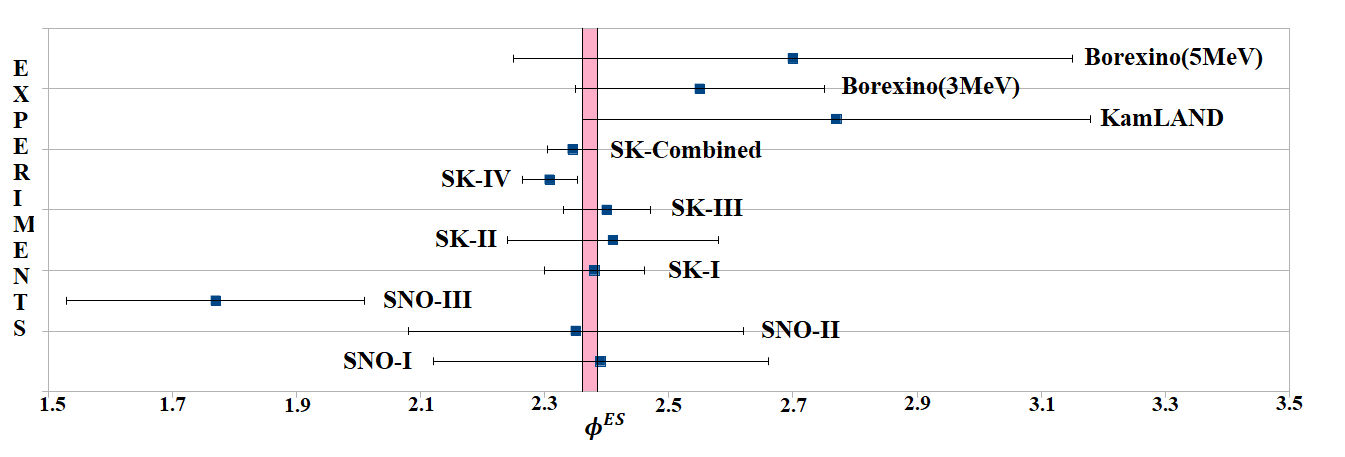}
\caption{$\phi^{ES}$ Overlap region (shaded) from all experiments.}
\label{figure:overlapflux}
\end{figure}

The SSM predictions \cite{SSM} for $^{8}$B solar neutrino flux to be observed at all experiments is taken to be $\phi_{SSM}=5.46 \pm 0.65 \times 10^{6}$ cm$^{-2}$s$^{-1}$. One has a number of possible combinations for doing  analysis, we use only the ones, which are giving us the most constrained limits on the unknown parameters under investigation. So, for first three analysis we choose the Sets in which we use fluxes $\phi^{NC},~\phi^{CC}$ from SNO-III with $\phi^{ES}$ from SNO-III, SK Combined, KamLAND, Borexino (3 MeV), Borexino (5 MeV), and flux of $\phi^{ES}$ common to all experiments as shown in Fig. {\ref{figure:overlapflux}}. We take all experiments one by one and derive constraints on f$_{B}$, active and sterile neutrinos. These Sets are shown in Table {\ref{Table:cases}}.\\

\begin{table}[H]
\centering
\begin{tabular}{|c|c|c|c|}
\hline 
\bf{Set} & \boldmath$\phi^{CC}$ & \boldmath$\phi^{NC}$ & \boldmath$\phi^{ES}$ \\ 
\hline 
1. & SNO-III & SNO-III & SNO-III \\ 
\hline 
2. & SNO-III & SNO-III & SK Combined \\ 
\hline 
3. & SNO-III & SNO-III & KamLAND \\ 
\hline 
4. & SNO-III & SNO-III & Borexino (3 MeV) \\ 
\hline 
5. & SNO-III & SNO-III & Borexino (5 MeV) \\ 
\hline 
6. & SNO-III & SNO-III & $\phi^{ES}$ Overlap \\ 
\hline 
\end{tabular} 
\caption{ Grouping of data in different sets.}
\label{Table:cases}
\end{table}

The values of r, $\bar{r}_{x}$, $\bar{r}_{d}$ used are shown in Table {\ref{Table:r}}:
\begin{table}[H]
\centering
\begin{tabular}{|c|c|c|c|}
\hline 
\bf{Experiment} & \bf{r} & \boldmath{$\bar{r}_{x}$} & \boldmath{$\bar{r}_{d}$} \\ 
\hline 
SNO-I & 0.150 & 0.115 & 0.954 \\ 
\hline 
SNO-II & 0.150 & 0.115 & 0.954 \\ 
\hline 
SNO-III & 0.151 & 0.116 & 0.955 \\ 
\hline 
SK-I & 0.149 & 0.114 & - \\ 
\hline 
SK-II & 0.151 & 0.116 & - \\ 
\hline 
SK-III & 0.151 & 0.116 & -\\ 
\hline 
SK-IV & 0.151 & 0.116 & - \\ 
\hline 
SK-Combined & 0.151 & 0.116 & - \\ 
\hline 
KamLAND & 0.210 & - & - \\ 
\hline 
Borexino (3 MeV) & 0.213 & 0.181 & - \\ 
\hline 
Borexino (5 MeV) & 0.213 & 0.181 & - \\ 
\hline 
\end{tabular} 
\caption{Cross-sectional ratios for various experiments .}
\label{Table:r}
\end{table}

\subsection{The sin$^{2}\alpha$ - f$_{B}$ Degeneracy}
\begin{figure}[H]
\begin{subfigure}[b]{0.4\textwidth}
\includegraphics[scale=0.6]{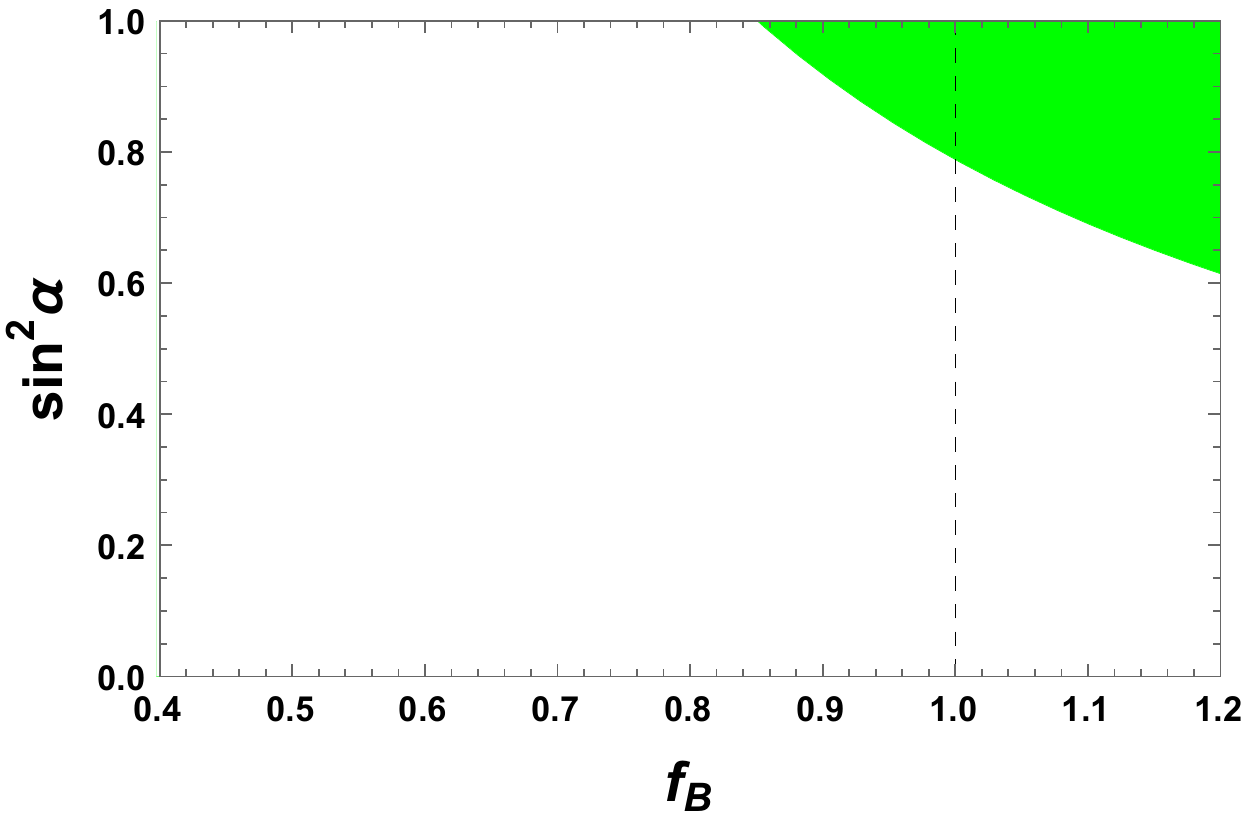}
\caption{sin$^2 \alpha$ vs f$_{B}$ plot for SNO-III.}
\end{subfigure}
\hspace{1cm}
\begin{subfigure}[b]{0.4\textwidth}
\includegraphics[scale=0.6]{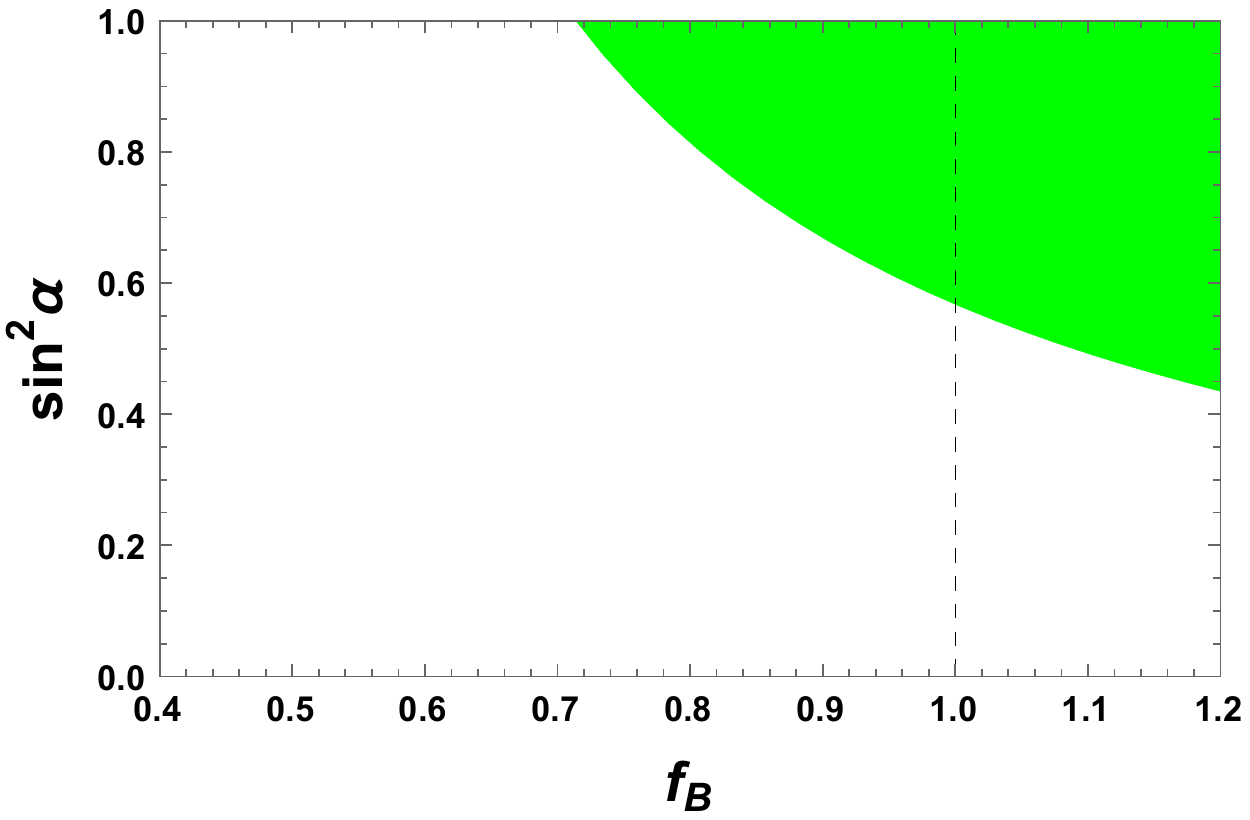}
\caption{sin$^2 \alpha$ vs f$_{B}$ plot for SK Comb.}
\end{subfigure}

\vspace{1cm}
\begin{subfigure}[b]{0.4\textwidth}
\includegraphics[scale=0.6]{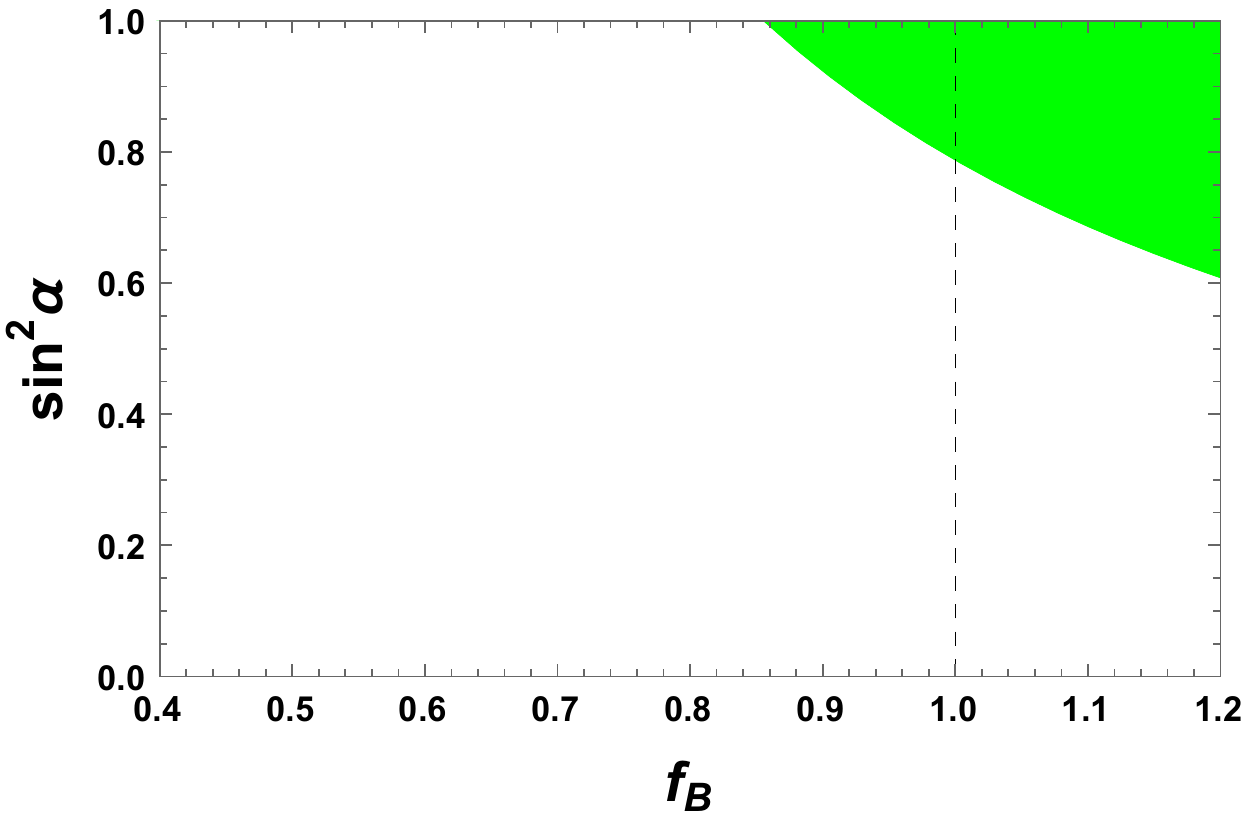}
\caption{sin$^2 \alpha$ vs f$_{B}$ plot for KamLAND.}
\end{subfigure}
\hspace{1cm}
\begin{subfigure}[b]{0.4\textwidth}
\includegraphics[scale=0.6]{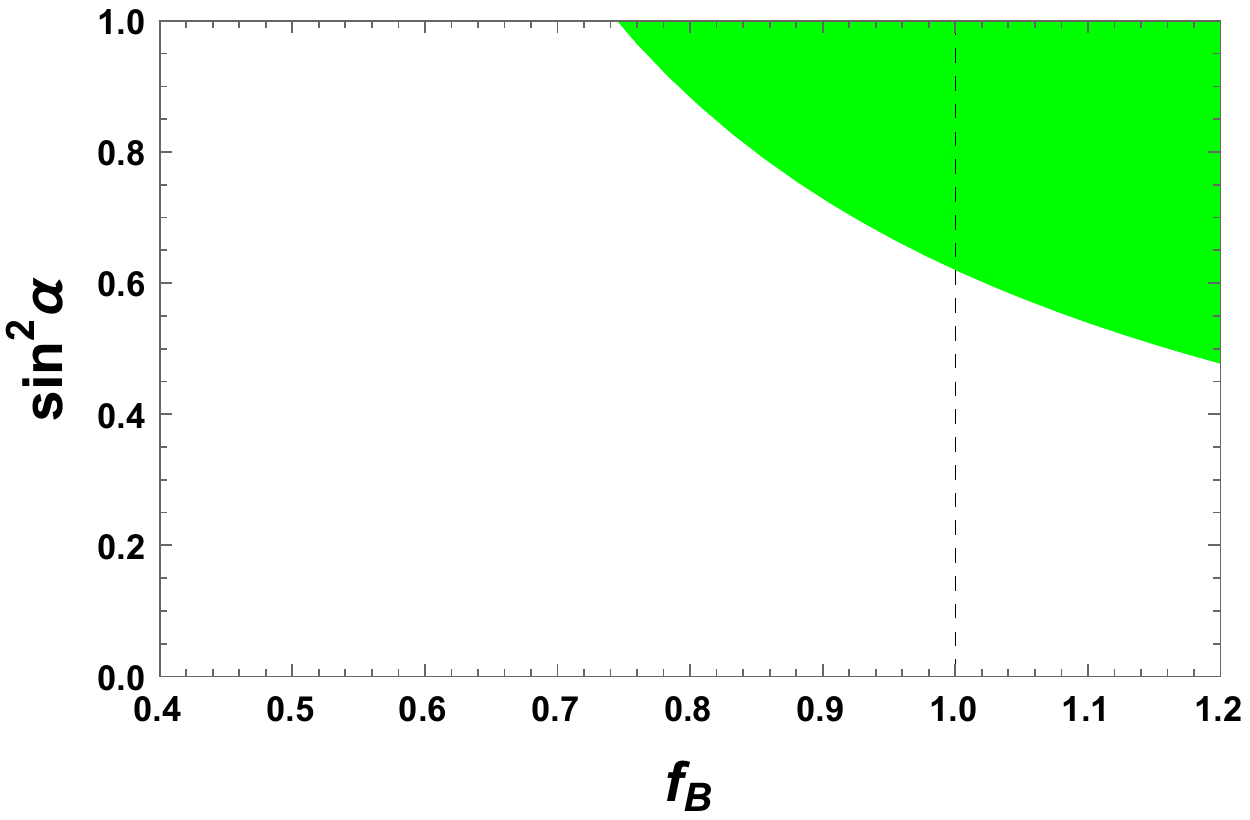}
\caption{sin$^2 \alpha$ vs f$_{B}$ plot for Borexino (3 MeV).}
\end{subfigure}

\vspace{1cm}
\begin{subfigure}[b]{0.4\textwidth}
\includegraphics[scale=0.6]{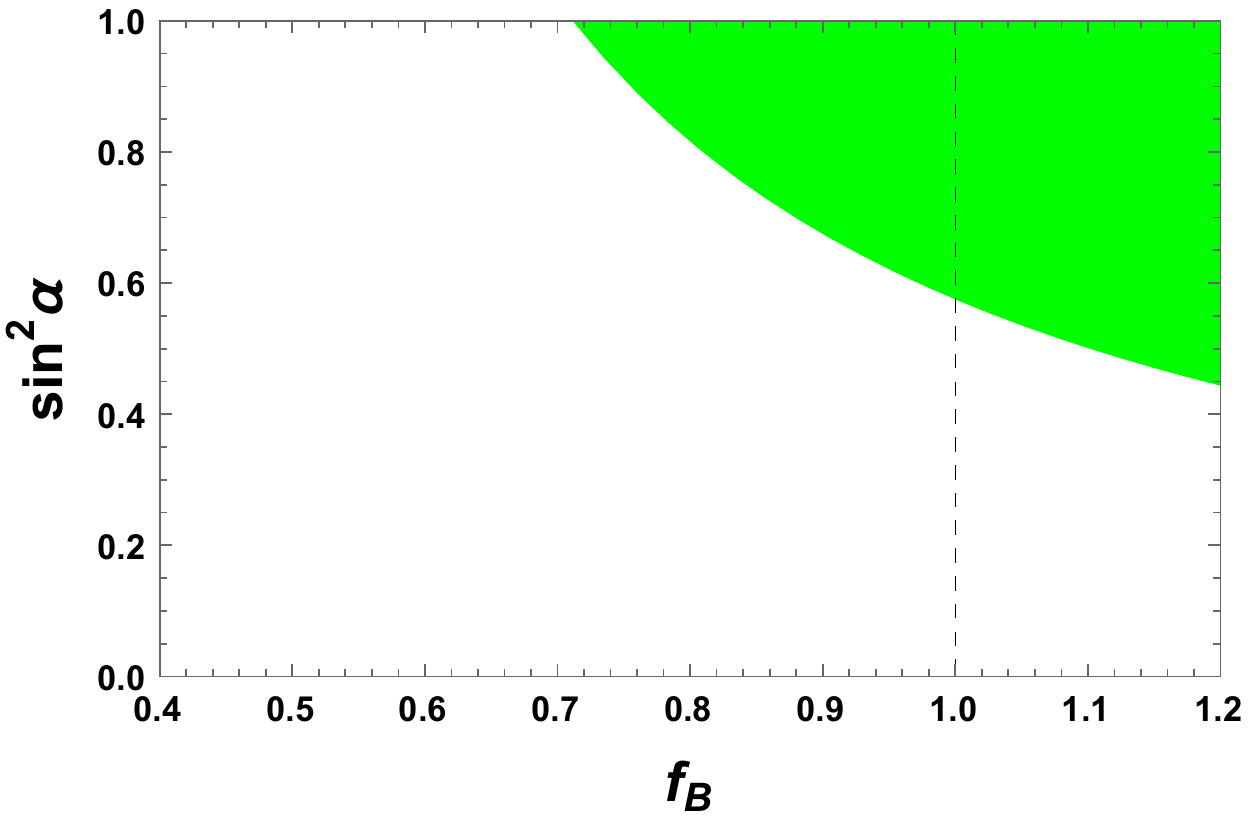}
\caption{sin$^2 \alpha$ vs f$_{B}$ plot for Borexino (5 MeV).}
\end{subfigure}
\hspace{1cm}
\begin{subfigure}[b]{0.4\textwidth}
\includegraphics[scale=0.6]{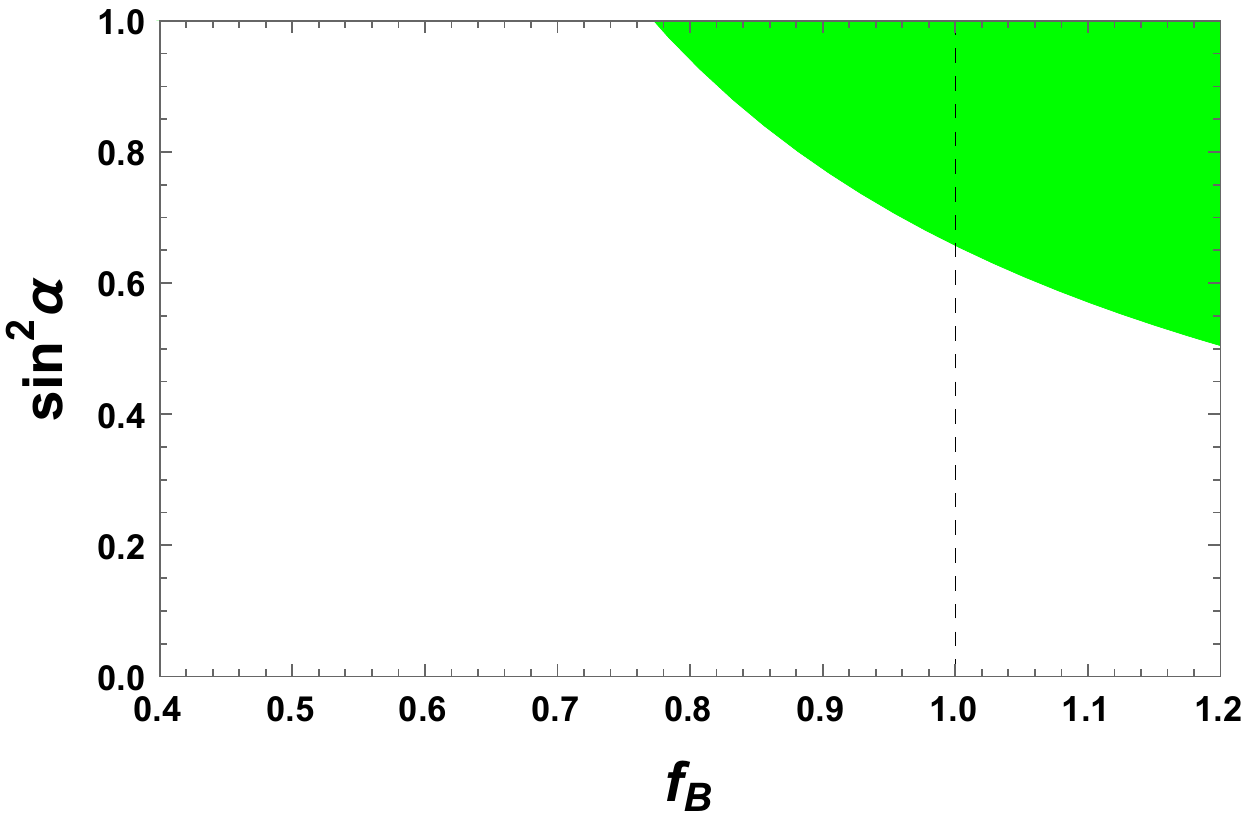}
\caption{sin$^2 \alpha$ vs f$_{B}$ plot for R$^{ES}$ Overlap.}
\end{subfigure}
\caption{sin$^2 \alpha$ vs f$_{B}$ common parameter space plots for all six Sets.}
\label{figure:1}
\end{figure}

 Using the equation which relates the f$_{B}$ and sin$^{2}\alpha$ i.e. equation \eqref{eq:19}, we obtain sin$^{2}\alpha$ - f$_{B}$ degeneracy plots for all the chosen Sets. Since we know from experiments that R$^{ES}_{i} >$ R$^{CC}_{SNO}$ (where i = SNO, SK, KamLAND and Borexino), possibility of sin$^2 \alpha = 0$ i.e. total conversion to sterile neutrinos, is rejected. However sin$^2 \alpha$ can approach to zero if f$_B$ is very large, which is unlikely. To make sin$^2 \alpha$ - f$_{B}$ plots, we vary f$_{B}$ from 0 to 1.2 and obtain a common parameter space. From this we get constraints on sterile neutrino percentage and the lower value of f$_{B}$. 
The variation of f$_{B}$ with sin$^2 \alpha$ for the fixed value of sin$^{2}{\psi}$ i.e. 1, obtained from the analysis is shown in the Fig. {\ref{figure:1}} for different Sets. Just to have reference, the dotted vertical line in the plots corresponds to the central value of f$_B = 1 \pm 0.12$\cite{SSM}.
 
 From these plots we calculate bounds on active neutrinos (sin$^2 \alpha $) and hence the sterile neutrino percentage as shown in Table \ref{Table:fb}.
 \begin{table}[H]
\centering
\begin{tabular}{|c|c|c|c|c|}
 \hline 
\bf {Set} & \bf {Data} & \bf {Bounds on f}\boldmath ${_B}$ & \bf {Bounds on sin}\boldmath ${^2 \alpha}$ & \bf {Sterile}\boldmath${\%}$ \\ 
 \hline 
1. & SNO-III & \boldmath {$0.85 \leq$ f$_B \leq 1.2$} & $0.613 \leq$ sin$^2 \alpha \leq 1$ & \boldmath{$0 - 38.7$} \\ 
 \hline 
 2. & SK Combined &  $0.71 \leq$ f$_B \leq  1.2$ & $0.434 \leq$ sin$^2 \alpha \leq 1$ & 0 - 56.6 \\ 
 \hline 
 3. & KamLAND &  \boldmath{$0.85 \leq$ f$_B \leq  1.2$} & $0.607 \leq$ sin$^2 \alpha \leq 1$ & \boldmath{$0 - 39.3$} \\ 
 \hline 
 4. & Borexino (3 MeV) &  $0.74 \leq$ f$_B \leq  1.2$ & $0.470 \leq$ sin$^2 \alpha \leq 1$ & 0 - 53.0 \\ 
 \hline 
 5. & Borexino (5 MeV) &  $0.71 \leq$ f$_B \leq  1.2$ & $0.443 \leq$ sin$^2 \alpha \leq 1$ & 0 - 55.7\\ 
 \hline 
 6. & R$^{ES}$ Overlap  &  $0.77 \leq$ f$_B \leq  1.2$ & $0.500 \leq$ sin$^2 \alpha \leq 1$ &0 - 50.0\\ 
 \hline 
 \end{tabular} 
 \caption{Results from sin$^2 \alpha$ - f$_{B}$ degeneracy plots.} 
 \label{Table:fb}
 \end{table}
 
 We find the most constrained results in case of SNO-III ($0-38.7\%$) and KamLAND ($0-39.3\%$) with 0.85 $\leq$ f$_B \leq 1.2 $, which is close to f$_B = 0.88$ i.e. the lower side of f$_B = 1 \pm$ 0.12\cite{SSM}. 

\subsection{ The $\phi$ - Active Flux}
We use the equations \eqref{eq:9}, \eqref{eq:12} and \eqref{eq:13} and obtain the results via two different analysis techniques - 1) sterile neutrino flux ($\phi_{sterile}$) and 2) sterile neutrino component percentage calculated as [(1-sin$^2 \alpha$)*100], i.e. depicted in column 6 of Table { \ref{Table:3}}. It is noted that, for no-sterile neutrinos, the CC/NC flux ratio in SNO is a direct measure of the average survival probability of $^{8}$B solar neutrinos that were detected through experiment as P$_{ee}=\phi^{CC}/\phi^{NC}$. Solving it with errors, we will have P$_{ee}$= 0.306 $\pm$ 0.04 with SSM and P$_{ee}$= 0.301 $\pm$ 0.03 with NC for high energy neutrinos. The difference in above two results for NC and SSM flux indicates the possibility for sterile neutrino flux present in the solar neutrino flux.

\begin{table}[H]
\centering
\begin{tabular}{|c|c|c|c|c|c|}
	\hline 
	\bf{Set} & \bf{Data Set} &\boldmath${\phi_{active}}$  & \boldmath$\phi_{sterile}$ & \bf {sin}\boldmath$^{2} \alpha $ & \bf{Sterile $\%$} \\  
	\hline
1. &	SNO-III & 6.313 $\pm$ 0.708 & -0.853 $\pm$ 0.961 & 1.225 $\pm$ 0.281 & \boldmath{$0 - 05.6$} \\
	\hline 
2. &	SK Combined & 5.395 $\pm$ 0.599  & 0.065 $\pm$ 0.884 & 0.983  $\pm$ 0.231 & 0 - 24.8   \\ 
	\hline 
3. &	KamLAND & 4.879 $\pm$ 1.185  & 0.581 $\pm$ 1.352  & 0.847 $\pm$ 0.345 & 0 - 49.8  \\ 
	\hline 
4. &	Borexino (3 MeV) & 5.428 $\pm$ 0.794   & 0.032 $\pm$ 1.026 & 0.991 $\pm$ 0.269 & 0 - 27.8 \\ 
	\hline 
5. &	Borexino (5 MeV)  & 5.127 $\pm$ 1.134  & 0.333 $\pm$ 1.307 & 0.912 $\pm$ 0.338 & 0 - 42.6 \\ 
	\hline 
6. &	$\phi^{ES}$ Overlap & 5.789 $\pm$ 0.686 & -0.329 $\pm$ 0.945 &   1.086 $\pm$ 0.259  & 0 - 27.3     \\
	\hline 
	
\end{tabular} 
    \caption{Constraints on active and  sterile neutrino fluxes in units of $10^6$ cm$^{-2}$s$^{-1}$.}
 \label{Table:3}
\end{table}
The constraints for active and sterile neutrino fluxes are depicted in columns 3 and 4 of Table {\ref{Table:3}}.
We calculate the upper bound on the  $\phi_{sterile}$ present in the solar neutrinos  for the all the Sets as:
 
\hspace{5cm} $\phi_{sterile} \leq 0.11 \times 10^6$ cm$^{-2}$s$^{-1}$ ( SNO-III)

\hspace{5cm} $\phi_{sterile} \leq 0.95 \times 10^6$ cm$^{-2}$s$^{-1}$ ( SK Combined )

\hspace{5cm} $\phi_{sterile} \leq 1.93 \times 10^6$ cm$^{-2}$s$^{-1}$ ( KamLAND)

\hspace{5cm} $\phi_{sterile} \leq 1.06 \times 10^6$ cm$^{-2}$s$^{-1}$ ( Borexino (3 MeV))

\hspace{5cm} $\phi_{sterile} \leq 1.64 \times 10^6$ cm$^{-2}$s$^{-1}$ ( Borexino (5 MeV))

\hspace{5cm} $\phi_{sterile} \leq 0.62 \times 10^6$ cm$^{-2}$s$^{-1}$ ($\phi^{ES}$ Overlap)

\vspace{.5cm}
 Column 5 of Table {\ref{Table:3}} represent the fraction of active neutrinos present in the solar neutrino flux using equation \eqref{eq:13}. The 1$\sigma$ range of sin$^{2}\alpha$ for KamLAND  indicates strong possibility, i.e. up to $49.8\%$, of sterile neutrino  (which is proportional to cos$^{2}\alpha$) percentage given in column 6 at 1$\sigma$ range. However, in the SK Combined case there is a possibility upto $24.8\%$  for sterile fraction at  1$\sigma$. The 1$\sigma$ range of sin$^{2}\alpha$ for Borexino (3 MeV) and Borexino (5 MeV) indicates much less possibility up to $27.8\%$ and $42.6\%$ respectively. However, in the Set-1 there is much less possibility for sterile fraction, which is $5.6\%$ at 1$\sigma$.


\subsection{The Medium Energy Neutrino Data}
\begin{figure}[H]

\centering
\includegraphics[scale=.8]{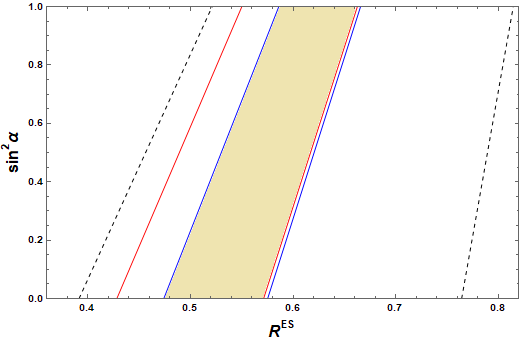}
\caption{1$\sigma $ plot of sin$^2 \alpha $ vs R$^{ES}$ for KamLAND (dotted lines), Borexino Phase-I (red lines) and Borexino Phase-II (blue lines). The shaded region corresponds to the overlapping region of all three experiments.} 
\label{figure:plot}
\end{figure}

Here, we use equation \eqref{eq:21}, which relates sin$^2 \alpha $ and R$^{ES}_{BOR/KL}$, and obtain sin$^2 \alpha $ vs R$^{ES}_{BOR/KL}$ plots for different values of P$_{ee}$ from different experiments. The values of P$_{ee}$ for different experiments is taken from Table {\ref{Table:medium}}. In Fig. {\ref{figure:plot}} we show a plot of sin$^2 \alpha $ vs R$^{ES}$ for KamLAND (dotted lines), Borexino Phase-I (red lines) and Borexino Phase-II (blue lines).

\begin{figure}[H]
\centering
\includegraphics[scale=.6]{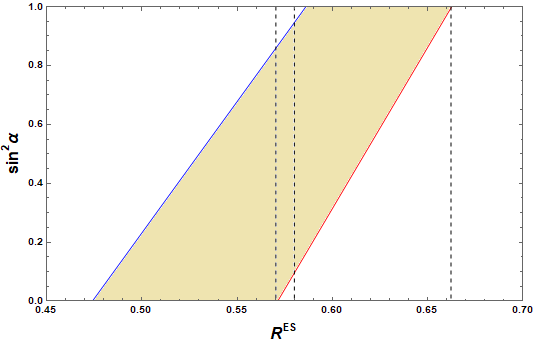}
\caption{$1\sigma $ plot of sin$^2 \alpha $ vs R$^{ES}$ zoomed for overlapping region of  KamLAND, Borexino Phase-I and Borexino Phase-II. Vertical dotted lines corresponds to R$^{ES}$ equal to $0.57$ , $0.58$ and $0.66$.} 
\label{figure:20}
\end{figure}

We zoom in the overlapped shaded region in Fig. \ref{figure:20} and project the three lines (dotted) vertically on it corresponding to three different arbitrary values of R$^{ES}$ to have some useful results. The fraction of sterile neutrinos is computed as 1 - sin$^2 \alpha$ (in the overlap region). It is clear from Fig. {\ref{figure:20}} that there is $14.03\%$ sterile neutrino component in the solar neutrino flux corresponding to R$^{ES}$ = 0.57 at $1\sigma$ and $15\%$ sterile neutrino component corresponding to R$^{ES}$ = 0.58 at $1\sigma$. Here R$^{ES} = 0.66$ corresponds to no sterile component.

 \subsection{\texorpdfstring{The $\chi^2$}{\chi\^{2}}-Analysis} \label{sec:5}
\begin{figure}[H]
\centering
\includegraphics[scale=.65]{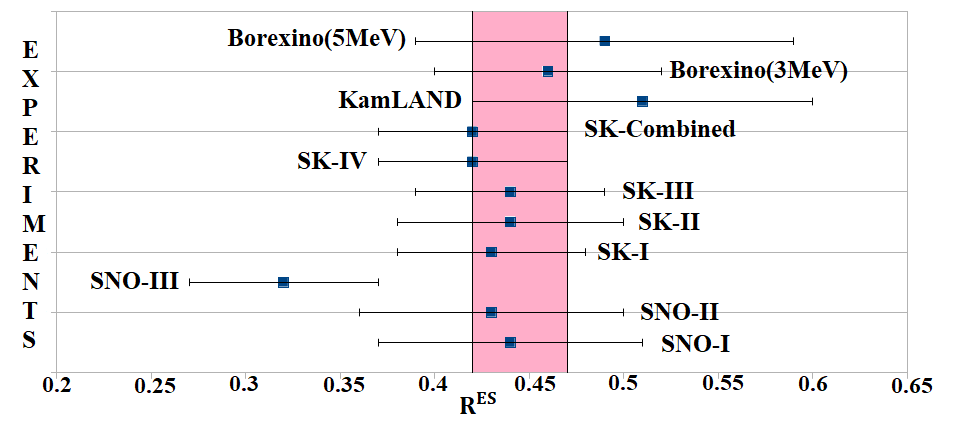}
\caption{ R$^{ES}$ Overlap region (shaded) from all experiments.}
\label{figure:2}
\end{figure}

In order to perform $\chi^2$-analysis, we focus on the reduced rates (henceforth to be called 'rates') of various experiments as shown in Table \ref{Table:2}. Therefore, the available data need to re-classified in various best possible cases to obtain useful results. In this exercise we arrange the data in various five cases as the most meaningful combinations. In Case-I we use R$^{NC}$, R$^{CC}$ and R$^{ES}$ from all the experiments. Then in Case-II we consider just the experimental results having best precision measurements. So we use R$^{NC}$ from SNO-II, R$^{CC}$ from SNO-I, SNO-II, SNO-III and R$^{ES}$ from SK-I, SK-III, SK-IV and SK Combined. In Case-III we use R$^{NC}$ and R$^{CC}$ from all the SNO-I, SNO-II, SNO-III experiments and R$^{ES}$ from all experiments except SNO-III as it has quite  different rate as compared to other experiments as shown in Fig. \ref{figure:2}. We consider the R$^{ES}$ of SNO-III as an independent case i.e. Case-IV, where we take R$^{NC}$ and R$^{CC}$ from all the SNO-I, SNO-II, SNO-III experiments and R$^{ES}$ from SNO-III only. In Case-V we use R$^{NC}$ and R$^{CC}$ from all the SNO-I, SNO-II, SNO-III experiments and R$^{ES}$ from the overlapping region of all the experiments for R$^{ES}$ as shown in Fig. {\ref{figure:2}}. All these different cases with various combinations of rates are shown in Table {\ref{Table:cases2}}.

Here, we perform the $\chi^{2}$-fitting for various cases with all the possible degrees of freedom using the $\chi^{2}$ formula given by the equation below:
\begin{equation}
\chi^{2}=\Sigma_{i}\frac{(R_{i}-R_{i}^{th})^{2}}{\delta R_{i}^{2}}.
\end{equation}

The sum extends over all the experimental data points (i $=$ ES$_{SK}$, ES$_{SNO}$, ES$_{KL}$, ES$_{Bor}$, NC, CC, etc.) as depicted in Table {\ref{Table:2}}.  Here R$_i$ and $\delta$R$_{i}$ denote the experimental rates and the corresponding errors, and R$^{th}_{i}$ are the theoretical values given by equations \eqref{eq:14}, \eqref{eq:15} and \eqref{eq:16}. $ \chi^2_{min} / n $  in column 5 of all the tables below is reduced $\chi ^2$-minimum, where $'n'$ represents the number of degrees of freedom (dof). We fix the value of sin$^2 \psi=1$, i.e no antineutrino component possibility as taken in the entire investigation.

Here, the maximum number of experimental data points is 17 (3R$^{NC}$+3R$^{CC}$+11R$^{ES}$) as can be seen from Table {\ref{Table:2}. However the maximum number of free parameters are three as shown in all the tables given below. As such, we get the number of dof accordingly for all the cases discussed.  
  
\begin{table}[H]
\centering
\def\arraystretch{1.5} \small

\begin{tabular}{|p{2cm}|p{2cm}|p{2cm}|p{7cm}|}

\hline 
\bf{Case} & \centering $\bf{R^{CC}}$ & \centering $\bf{R^{NC}}$ & $\bf{R^{ES}}$  \\
\hline 
I & SNO-I, SNO-II, SNO-III. & SNO-I, SNO-II, SNO-III. & SNO-I, SNO-II, SNO-III, SK-I, SK-II, SK-III, SK-IV, SK Combined,
KamLAND, Borexino (3 MeV) and Borexino (5 MeV). \\
  
\hline 
II & SNO-I, SNO-II, SNO-III. & SNO-II & SK-I, SK-III, SK-IV and SK Combined. \\ 
\hline 
III & SNO-I, SNO-II, SNO-III. & SNO-I, SNO-II, SNO-III. &  SNO-I, SNO-II, SK-I, SK-II, SK-III, SK-IV, SK Combined,
KamLAND, Borexino (3 MeV) and Borexino (5 MeV). \\ 
\hline 
IV & SNO-I, SNO-II, SNO-III. & SNO-I, SNO-II, SNO-III. & SNO-III \\ 
\hline 
V & SNO-I, SNO-II, SNO-III. & SNO-I, SNO-II, SNO-III. &  R$^{ES}$ Overlap \\ 
\hline 
\end{tabular} 
\caption{Different Cases for $\chi ^2$-analysis.}
\label{Table:cases2}
\end{table}

\begin{table}[H]
\begin{center}
\begin{tabular}{|c|c|c|c|c|c|c|}
\hline 
 \bf{Case} & \bf{No. of dof} & $\bf{f_{B}}$ & $\bf{P_{ee}}$ & \bf{sin}\boldmath{$^2 \alpha$} & \boldmath$\frac {\chi^2_{min}}{n}\ $ & \bf{ Sterile} \boldmath $\%$ ($1\sigma$)\\ 
\hline  
& 16  & $0.974$ & $0.327$ & $0.970$ & $0.363$ & 0 - 31.5 \\ 
I & 15  & $0.975$ & $0.327$ & $0.970$ & $0.387$ & 0 - 31.5 \\ 
& 14  & $0.975$ & $0.327$ & $0.970$ & $0.415$ & 0 - 31.5 \\ 
\hline 
\hline  
& 7  & $0.962$ & $0.332$ & $0.970$ & $0.138$ & 0 - 22.5 \\ 
II & 6  & $0.965$ & $0.332$ & $0.970$ & $0.161$ & 0 - 22.5 \\ 
& 5  & $0.965$ & $0.332$ & $0.970$ & $0.194$ & 0 - 22.5 \\ 
\hline 
\hline 
& 15 & $0.982$ & $0.330$ & $0.970$ & $0.127$ & 0 - 26.8 \\ 
III & 14  & $0.985$ & $0.330$ & $0.970$ & $0.136$ & 0 - 26.8 \\ 
& 13  & $0.985$ & $0.330$ & $0.970$ & $0.146$ & 0 - 26.8 \\ 
\hline 
\hline  
& 6  & $0.921$ & $0.322$ & $0.970$ & $0.441$ & 0 - 19.4 \\ 
IV & 5  & $0.924$ & $0.322$ & $0.970$ & $0.530$ & 0 - 19.4 \\ 
& 4  & $0.924$ & $0.322$ & $0.970$ & $0.662$ & 0 - 19.4 \\ 
\hline 
\hline  
& 6  & $0.956$ & $0.328$ & $0.970$ & $0.118$ & 0 - 18.4 \\ 
V & 5  & $0.957$ & $0.328$ & $0.970$ & $0.141$ & 0 - 18.4 \\ 
& 4  & $0.957$ & $0.328$ & $0.970$ & $0.177$ & 0 - 18.4 \\ 
\hline

\end{tabular} 
\end{center}
\caption{Results of $\chi^2$-analysis for all cases.}
\label{Table:6}
\end{table}

\begin{figure}[H]

\begin{subfigure}[b]{0.45\textwidth}
\includegraphics[width=8cm, height=6cm]{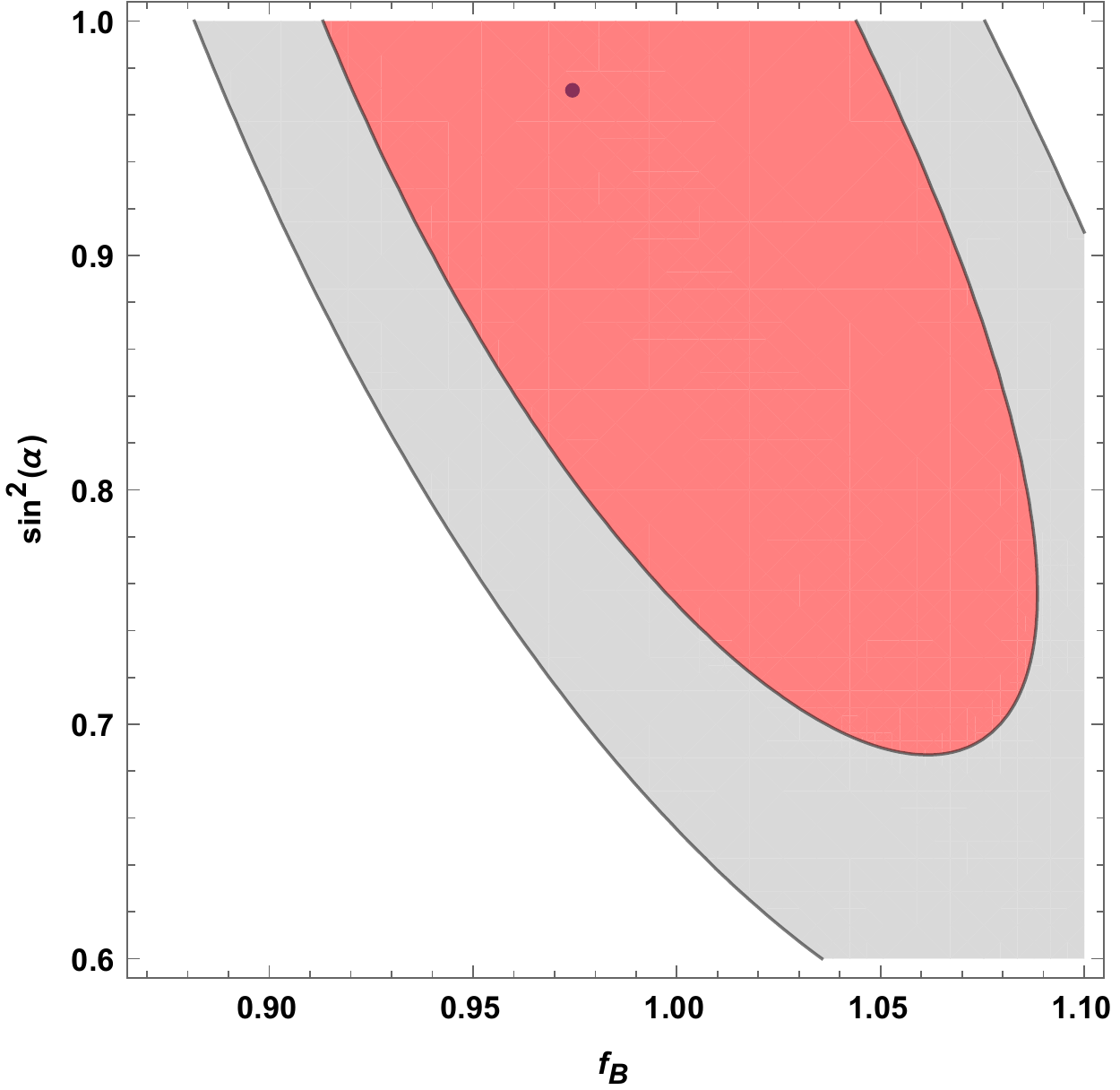}
\caption{1$\sigma$, 2$\sigma$ contour plots in the f$_B$, sin$^2{\alpha}$ plane for Case-I.            }
\end{subfigure}
~\qquad
\hspace{.4cm}
\begin{subfigure}[b]{0.45\textwidth}
\includegraphics[width=8cm, height=6cm]{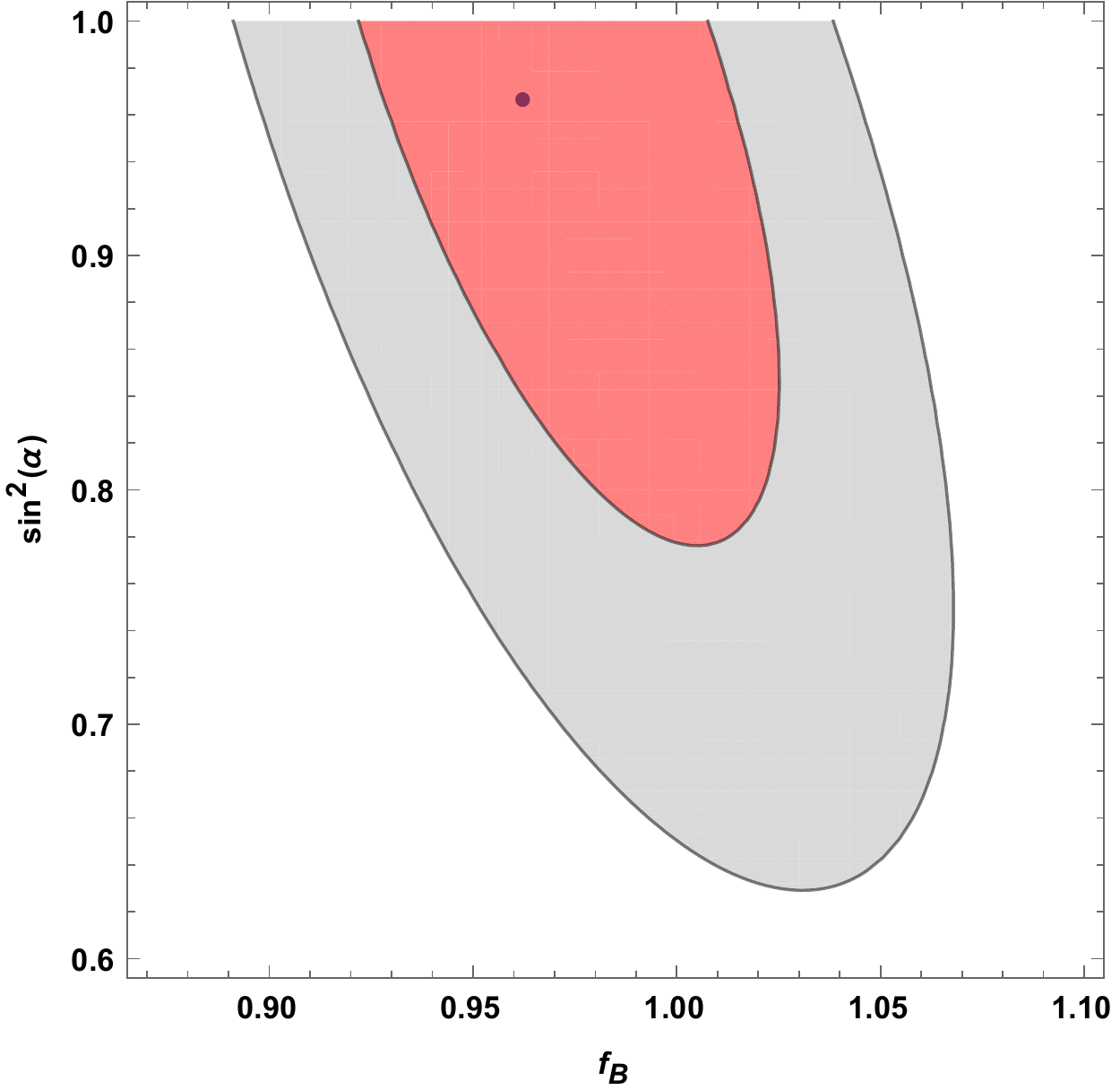}
\caption{1$\sigma$, 2$\sigma$ contour plots in the f$_B$, sin$^2{\alpha}$ plane for Case-II.}
\end{subfigure}
~\qquad

\vspace{0.3cm}
\begin{subfigure}[b]{0.45\textwidth}
\includegraphics[width=8cm, height=6cm]{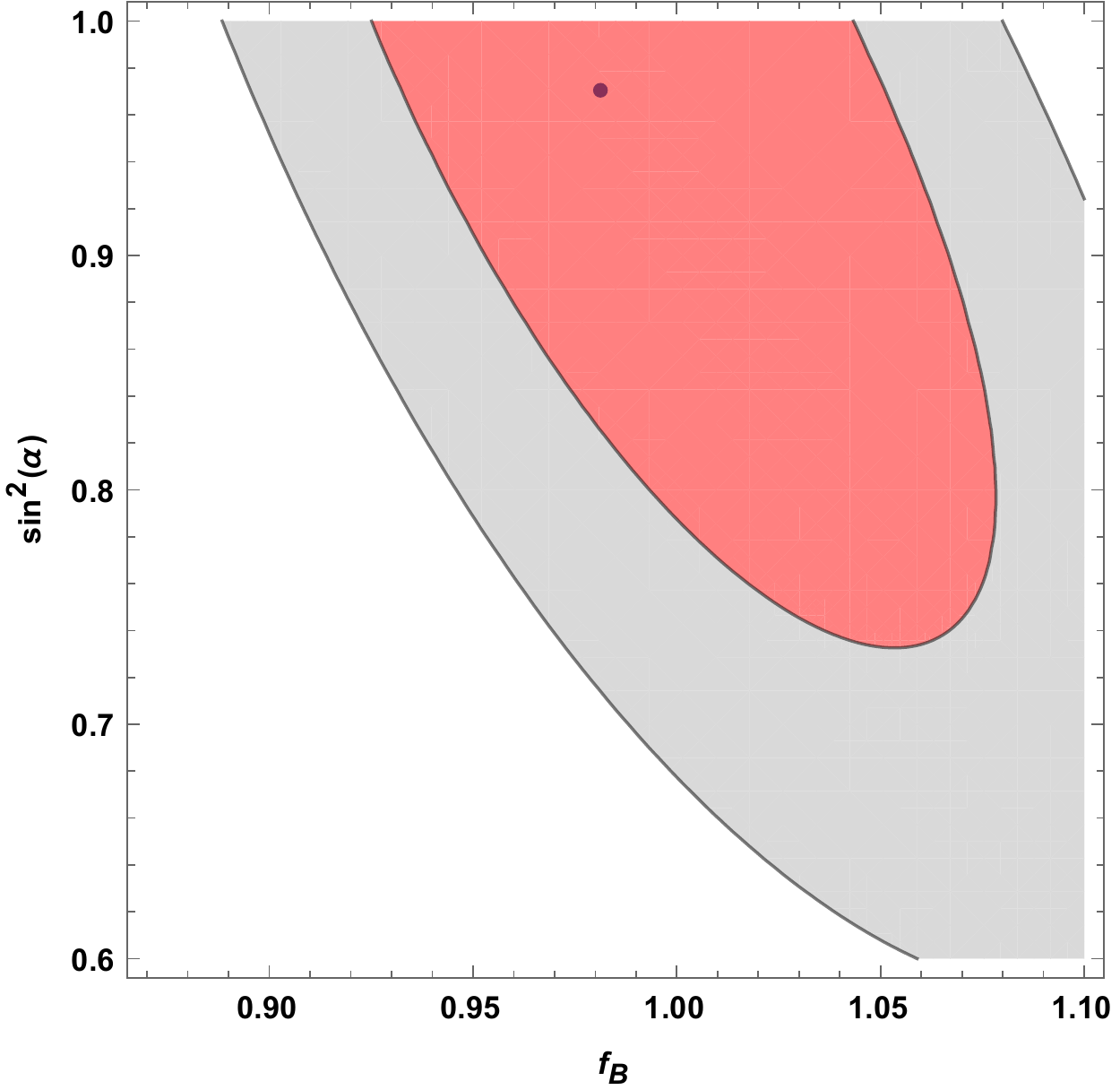}
\caption{1$\sigma$, 2$\sigma$ contour plots in the f$_B$, sin$^2{\alpha}$ plane for Case-III.}
\end{subfigure}
~\qquad
\hspace{.4cm}
\begin{subfigure}[b]{0.45\textwidth}
\includegraphics[width=8cm, height=6cm]{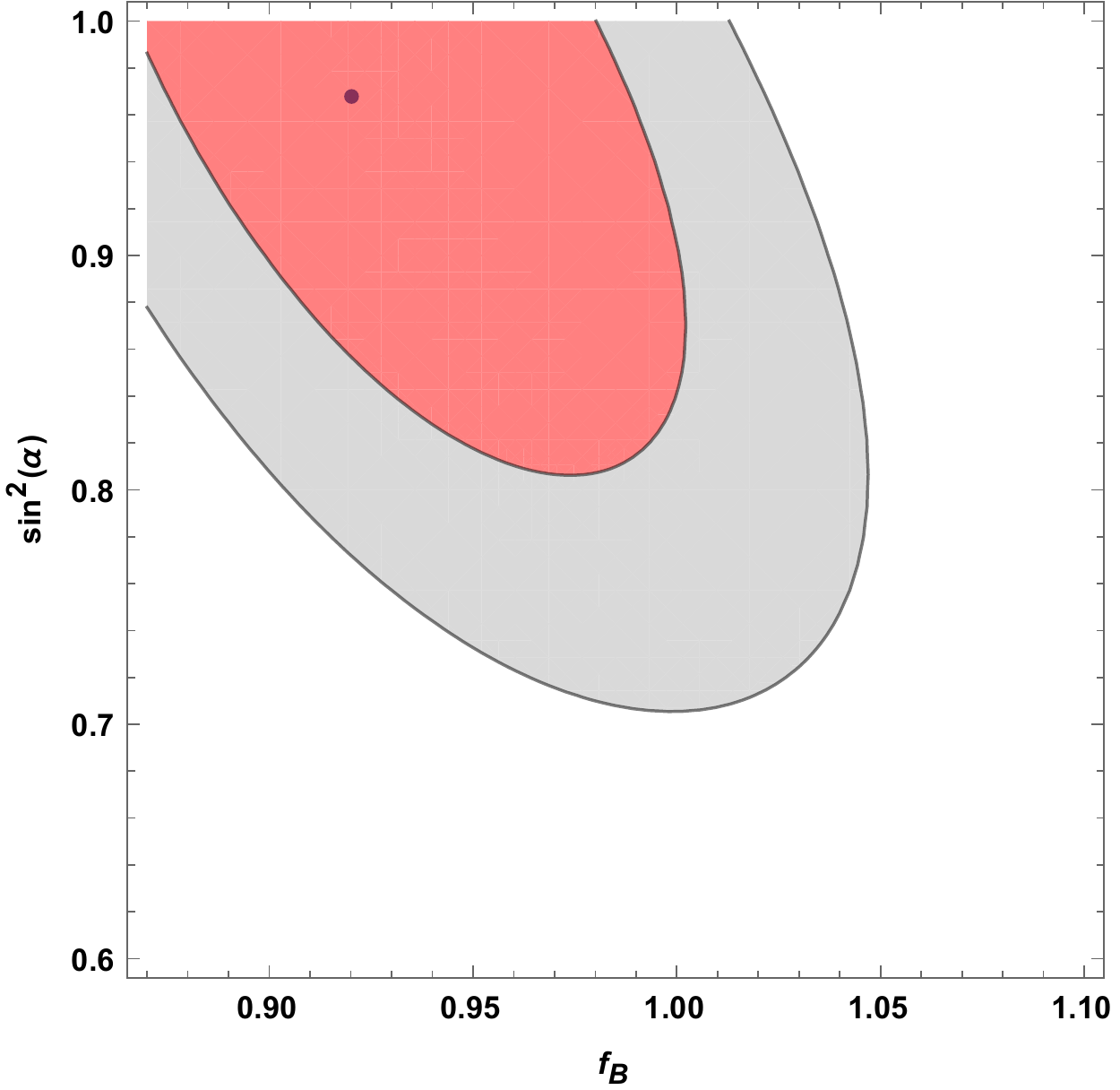}
\caption{1$\sigma$, 2$\sigma$ contour plots in the f$_B$, sin$^2{\alpha}$ plane for Case-IV.}
\end{subfigure}
~\qquad

\vspace{0.3cm}
\centering
\begin{subfigure}[b]{0.6\textwidth}
\centering
\includegraphics[width=8cm, height=6cm]{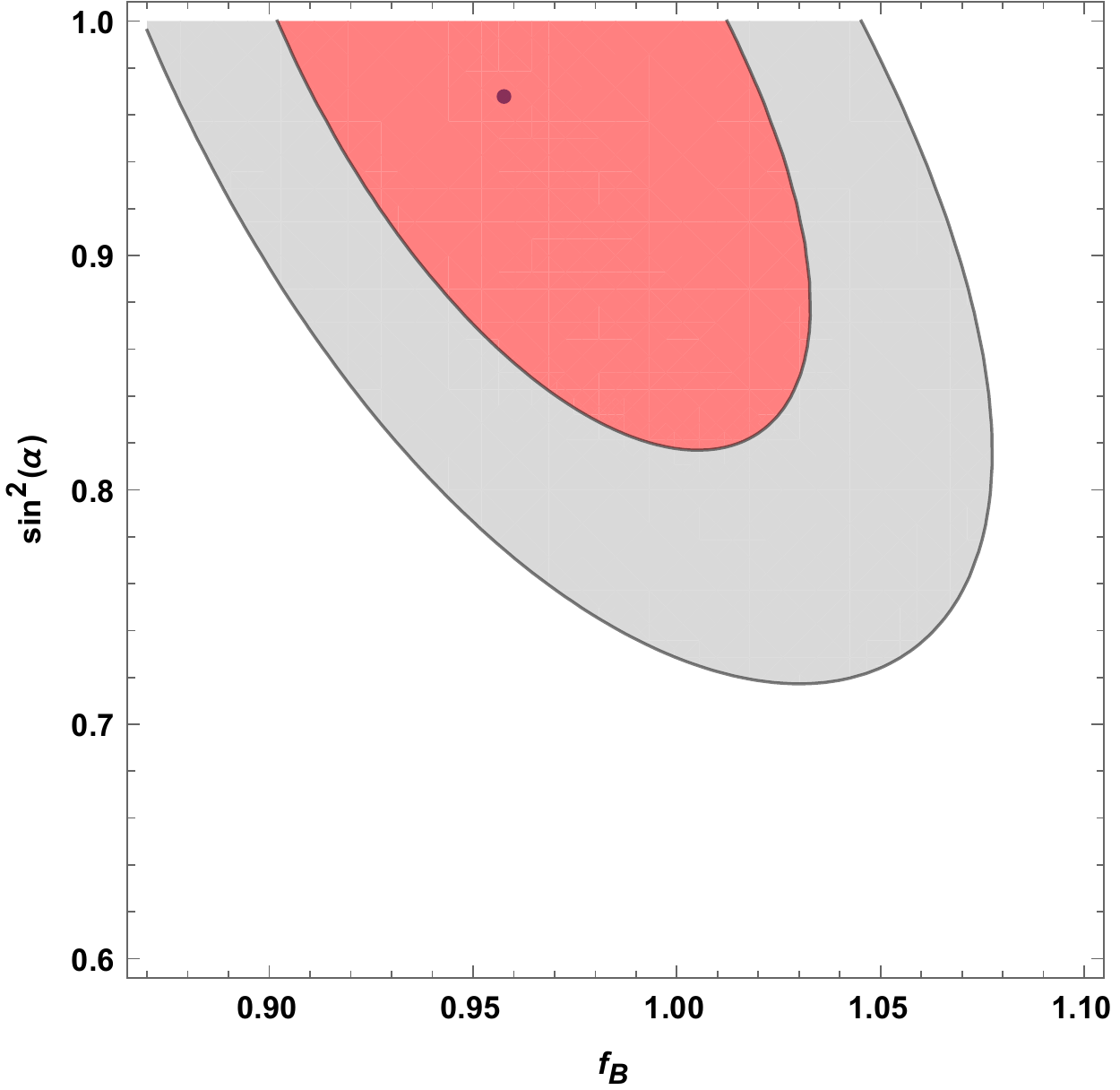}
\caption{1$\sigma$, 2$\sigma$ contour plots in the f$_B$, sin$^2{\alpha}$ plane for Case-V.}

\end{subfigure}
\caption{1$\sigma$, 2$\sigma$ contour plots for all five Cases. The dot represents the best fit point.}
\label{figure:3}
\end{figure}

From all these plots and $\chi^2$-fitting tables, it is interesting to note that the best fit point for active neutrino fraction gives incidentally the same value for all the cases as shown in the tables above, and i.e. sin$^2 \alpha = 0.970$. This leads to just $3\%$ sterile neutrino fraction given by all the cases irrespective of the data combinations taken. 

 As can be seen from Fig. {\ref{figure:3}}, the most constrained results correspond to Case-V. In this case, there is the least fraction of sterile neutrino  component which is $18.4\%$ at $1\sigma$, as shown in the last row of Table {\ref{Table:6}}. In the same Table {\ref{Table:6}}, the sterile neutrino fractions for all other cases at  $1\sigma$ are also shown. For Case-I, a possibility of $31.5\%$, for Case-II $22.5\%$ and, $26.8\%$ and $19.4\%$ sterile neutrino fractions are shown for Case-III and Case-IV, respectively.

In all the above four subsections, we discuss five different model independent analysis techniques to constrain the percentage and flux of sterile neutrinos. From  the results of the four analysis techniques, we get sterile neutrino percentages summarized in Table \ref{Table:comparision}. 
\begin{table}[h]
\begin{tabular}{|c|c|c|c|c|}
\hline 
\bf{S.No.} &  \bf{sin}\boldmath{$^2 \alpha$ }\bf{- f}\boldmath{$_B$} \bf Degeneracy & \bf{sin}\boldmath{$^2 \alpha$} & \bf{Medium Energy Analysis} & \boldmath$ {\chi^2}$ \bf {- Analysis} \\ 
\hline 
1 &  \boldmath{$0 - 38.7 \%$} & \boldmath{$0 - 05.6 \%$} & \boldmath{$0 - 14.03 \%$} & $0 - 31.5 \%$ \\ 
\hline 
2 & $0 - 56.6 \%$ & $0 - 24.8 \%$ & $0 -  15 \%$ & $0 - 22.5 \%$ \\ 
\hline 
3 & \boldmath{$0 - 39.3 \%$} & $0 - 49.8 \%$ & - & $0 - 26.8 \%$ \\ 
\hline 
4 & $0 - 53.0 \%$ & $0 - 27.8 \%$ & - & $0 - 19.4 \%$ \\ 
\hline 
5 & $0 - 55.7 \%$ & $0 - 42.6 \%$ & - & \boldmath{$0 - 18.4 \%$} \\ 
\hline 
6 & $0 - 50.0 \%$ & $0 - 27.3 \%$ & - & - \\ 
\hline 
\end{tabular} 
\caption{Sterile neutrino percentage obtained from various model independent ways.}
\label{Table:comparision}
\end{table}

 We can see that from sin$^2 \alpha$ - f$_B$ degeneracy, most constrained limits on sterile neutrino component at $1\sigma $ are obtained in Set-1 ($38.7 \%$) and Set-3 ($39.3 \%$), where we use SNO-III and KamLAND data respectively. From sin$^2 \alpha $, we see that in the Set-1, there is much less possibility for sterile fraction, which is $5.6\%$ at $1\sigma$. In these two techniques shown in columns 2 and 3 in Table \ref{Table:comparision}, we  use same Sets which are described earlier. Medium energy analysis gives a sterile neutrino percentage of $0-15 \% $. From  $\chi^2 $ analysis we obtain most constrained limit as $18.4 \% $ in Case-V as shown in Table \ref{Table:comparision}.
However, the fifth technique puts constraints on the fluxes of sterile neutrinos as shown in subsection 4.2. It is clear that more constrained upper bounds on $\phi_{sterile}$ are obtained in Set-1 (SNO-III). It has been noted that the Set-6, in which we use $\phi^{ES}$ Overlap is also giving quite constrained bounds on sterile neutrino flux.

\section{Conclusions} \label{sec:6}
The constraints on the flux of sterile neutrinos present in solar neutrino data would reveal many riddles still hidden in the solar neutrino physics, as well as in the neutrino astrophysics as a whole. In this work, we derive such constraints in a model independent way i.e. on the flux of sterile neutrinos $\nu_s$, which are more stringent as compared to the existing in literature. The global and comprehensive analysis of the solar neutrino data available from all the solar neutrino experiments including KamLAND solar phase and Borexino (3 MeV), Borexino (5 MeV) have been done. This kind of updated analysis has been done for the first time, and thereby appearing in literature. Therefore, it becomes interesting to present the bounds obtained by such investigation. 

From sin$^2 \alpha$ - f$_{B}$ degeneracy plots, it is clear that most constrained limits on sterile component are obtained in Set-1 (38.7 $\%$) and Set-3 ($39.3 \%$), where we use SNO-III and KamLAND data respectively. It is clear that more constrained upper bounds on $\phi_{sterile}$ are obtained in Set-1 (SNO-III). Set-6 in which we are using $\phi^{ES}$ Overlap is also giving very constrained bounds on sterile neutrino flux. It may be noted that these bounds are more constrained as compared to the ones obtained earlier in literature \cite{Barger:2001pf}, \cite{bmw}, \cite{kang}, \cite{bcc}, \cite{skumar} and \cite{lalsingh1}. In Table {\ref{Table:3}}, we observe that in Set-1, there is much less possibility for sterile neutrino fraction, which is $5.6\%$ at $1\sigma$. From medium energy data analysis, we see that there is a possibility of no sterile component corresponding to reduced R$^{ES} = 0.66$. Also from $\chi^2 $ analysis we obtain that there is a posssibility of $31.5\%$, $22.5\%$, $26.8\%$, $19.4\%$ and $18.4\%$ sterile component for Case-I, II, III, IV and V respectively as shown in Table {\ref{Table:6}}.


The indication for the existence of sterile neutrinos has emerged from Wilkinson Microwave Anisotropy Probe (WMAP), data as well, which suggested that the number of neutrino families in the early universe was more than three \cite{WMAP}. There are several models suggested in literature, which fit the sterile neutrinos in the scheme like a 3+1 model i.e. the three ordinary neutrinos and a sterile one. From the present work it is clear that there is a substantial number of sterile neutrinos and active-sterile mixing in solar neutrino flux. Such additional mass eigen states are light and could be of $\sim$1 eV scale as assumed in 3+1 and other particle physics and cosmological models including light sterile neutrinos and their mixing.  

The Liquid Scintillator Neutrino Detector (LSND) anomaly indicated the signal of light sterile neutrinos of mass scale $\sim$1 eV. The data from Mini Booster Neutrino Experiment (MiniBooNE) are consistent in energy and magnitude with the excess of events reported by the LSND, and this excess is significant at 6.0$\sigma$, when the results of both experiments are combined together \cite{LSND}. On the other hand, the joint analysis of Main Injector Neutrino Oscillation Search (MINOS) and MINOS+ (continuation of MINOS) experiments sets stringent limits on active-sterile mixing (in 3+1 model) for values of $\Delta$m$^2_{41}$ >  10$^{-2}$ eV$^2$. However, the final year of MINOS+ data with ongoing analysis improvements will increase the senstivity for future limits on sterile neutrinos even further \cite{MINOS}. As such, more statistics and analysis techniques are awaited to conclude about the issue. Sterile neutrinos are well motivated extensions of the SM, and symmetry protected seesaw scenarios allow for  electroweak scale sterile neutrino masses and active-sterile mixings. This work strongly indicates the existence of sterile neutrinos and therefore, such a study is quite promising for the future searches of this class of neutrinos.

\vspace{.6cm}
\begin{large}
\textbf{Acknowledgements}
\end{large} \\
 We acknowledge the financial support provided by the UGC, Govt. of India under the project grant MRP-MAJOR-PHYS-2013-12281. One of the authors BCC thanks
IUCAA for the providing hospitality during the preparation of the work. The earlier version of this manuscript has been presented as conference abstract in "International Conference on Sciences : Emerging Scenario and Future Challenges (SESFC-2016)"\cite{gazal}. We thank Lal Singh for useful discussion.


\begin{thebibliography}{99}
\bibitem{Davis:1968cp}
R.~J.~Davis, D.~S.~Harmer and K.~C.~Hoffman,
Phys.~Rev.~Lett.~{\bf 20}, 1205 (1968).
\bibitem{Cleve:1998}
B.~T.~Cleveland {\it {et al.}} [Homestake Collaboration], 
Astrophys. J. {\bf 496}, 505 (1998).
\bibitem{Eguchi:2002dm}
K.~Eguchi {\it et al.}  [KamLAND Collaboration],
Phys.~Rev.~Lett.~{\bf 90}, 021802 (2003).
\bibitem{LMA} 
P.~Aliani, V.~Antonelli, R.~Ferrari, M.~Picariello and E.~Torrente-Lujan,
arXiv:0406182[hep-ph].
\bibitem{LMA1}
A.B. Balantekin, H. Yuksel, Phys. Rev. {\bf D68}, 113002 (2003). 
\bibitem{LMA2} 
P.~Aliani, V.~Antonelli, M.~Picariello and E.~Torrente-Lujan,
New J.~Phys.~{\bf 5}, 2 (2003). 
\bibitem{LMA3}
M. Maltoni, T. Schwetz, M.A. Tortola, J.W.F. Valle,  Phys.Rev.
{\bf D68}, 113010 (2003).
\bibitem{LMA4}
 A. Bandyopadhyay,
S. Choubey, S. Goswami, S.T. Petcov, D.P. Roy, Phys. Lett. {\bf B583}, 134 (2004).
\bibitem{LMA5}
 P. Aliani, V. Antonelli, M. Picariello, E. Torrente-Lujan, 
Phys. Rev. {\bf D69}, 013005 (2004).
\bibitem{LMA6}
P. C. de Holanda and A. Yu. Smirnov, Astropart. Phys {\bf 21}, 287-301 (2004).
\bibitem{LMA7} 
J. N. Bahcall, M. C. Gonzalez-Garcia, C. Pe\~{n}a-Garay,  JHEP {\bf 0302}, 009 (2003).

\bibitem{sterile}
B. C. Chauhan and J. Pulido, JHEP {\bf 0406}, 008 (2004).
\bibitem{sterile1}
 A. Yu. Smirnov, Nucl. Phys. Proc. Suppl. {\bf235}, 431 (2013).
 \bibitem{sterile2}
  L. B. Auerbach {\it et al.} [LSND Collaboration], Phys. Rev. {\bf C64}, 065501 (2001).
 \bibitem{sterile3}
  M. Antonello {\it et al.}, Eur. Phys. J. {\bf C73}, 2345 (2013).
  \bibitem{sterile4}
   P. C. de Holanda and A. Yu. Smirnov, Phys. Rev. {\bf D83}, 113011 (2011).
\bibitem{Sturrock:2004hx}
P.~A.~Sturrock, D.~O.~Caldwell, J.~D.~Scargle, G.~Walther and M.~S.~Wheatland,
arXiv:0403246[hep-ph].
\bibitem{Sturrock:2003kv}
P.~A.~Sturrock,
Astrophys.~J.~{\bf 605}, 568 (2004).
\bibitem{Caldwell:2003dw}
D.~O.~Caldwell and P.~A.~Sturrock,
arXiv:0309191[hep-ph].
\bibitem{Sturrock:2001qn}
P.~A.~Sturrock and M.~A.~Weber,
Astrophys.~J.~{\bf 565}, 1366 (2002).
\bibitem{Berezinsky:2002fa}
V.~Berezinsky, M.~Narayan and F.~Vissani,
Nucl.~Phys.~{\bf B658}, 254 (2003).
\bibitem{deHolanda:2002ma}
P.~C.~de Holanda and A.~Y.~Smirnov, Phy. Rev. {\bf D69}, 113002 (2004).


\bibitem{Eguchi:2003gg}
K.~Eguchi {\it et al.} [KamLAND Collaboration],
Phys. Rev. Lett. {\bf 92}, 071301 (2004).
\bibitem{sno} 
B. Aharmim {\it{et al.}} [SNO Collaboration], Phys. Rev. {\bf C87}, 015502 (2013).

\bibitem{ncd}
B. Abharmim {\it {et al.}} [SNO Collaboration],
Phys. Rev. Lett. {\bf101}, 111301 (2008).
\bibitem{sk}
K. Abe {\it {et al.}} [The Super-Kamiokande Collaboration], 
Phys. Rev. {\bf D94}, 052010 (2016).
\bibitem{BorII}
M. Agostini {\it{et al.}} [Borexino Collaboration], arXiv:1709.00756v1[hep-ex]. 
\bibitem{Bor}
G. Bellini {\it{et al.}} [Borexino Collaboration], Phys. Rev. {\bf D82}, 033006 (2010).

\bibitem{kl}
S. Abe {\it{et al.}} [The KamLAND Collaboration], Phys. Rev. {\bf C84}, 035804 (2011). 
\bibitem{Barger:2001pf}
V.~D.~Barger, D.~Marfatia and K.~Whisnant,
Phys.~Lett.~{\bf B509}, 19 (2001).
\bibitem{Barger:2001zs}
V.~D.~Barger, D.~Marfatia and K.~Whisnant,
Phys.~Rev.~Lett. {\bf 88}, 011302 (2002).
\bibitem{Barger:2002iv}
V.~Barger, D.~Marfatia, K.~Whisnant and B.~P.~Wood,
Phys.~Lett.~{\bf B537}, 179 (2002).
\bibitem{Balantekin:2004hi}
A.~B.~Balantekin, V.~Barger, D.~Marfatia, S.~Pakvasa and H.~Yuksel,
Phy. Lett. {\bf B613}, 61 (2005).
\bibitem{bmw} V. Barger, {\it et al}. Phys. Rev. Lett. {\bf 88}, 011302 (2002); Phys. Lett. {\bf B537},  179 (2002).
\bibitem{Band} 
A. Bandyopadhyay, {\it et al}. Phys. Lett. {\bf B540}, 14 (2002).
\bibitem{kang} S. K. Kang and C. S. Kim, Phys. Lett. {\bf B584}, 98-102 (2004).
\bibitem{bcc}
Bhag~C.~Chauhan and J.~Pulido,
 JHEP{\bf 0412}, 040 (2004).
\bibitem{skumar} S. Dev, Sanjeev Kumar and Surender Verma, Modern Physics Letters {\bf A21}, 1761 (2006).
\bibitem{lalsingh1}
 Lal Singh, Bhag C. Chauhan, Ravi Dutt, K. K. Sharma and S. Dev, arXiv:1102.4917v1 [hep-ph].
\bibitem{sk4} H. Nishino {\it et al.} Null. Instrum. Methods Pays. Res. Sect. {\bf A
610}, 710 (2010).
\bibitem{Yamada}
 S. Yamada {\it et al.}, IEEE Trans. Nucl. Sci. {\bf57}, 428 (2010).
\bibitem{bekam}
A. Gando {\it et al.} [KamLAND Collaboration], Phys. Rev. {\bf C92}, 055808 (2015).

\bibitem{SSM}
Nuria Vinyoles {\it {et al.}} The Astrophysical Journal, 
{\bf 835}, 2 (2017).
 \bibitem{bebor2}
M.Agostini {\it et al.} [Borexino Collaboration], arXiv:1707.09279v2[hep-ex].




\bibitem{WMAP} Komatsu {\it et al.} [WMAP Collaboration], Astrophys. J. Suppl., 192 (2011).
\bibitem{LSND}
A. A. Aguilar-Arevalo {\it et al.} [MiniBooNE Collaboration], Phys. Rev. Lett. {\bf121}, 221801 (2008).
 
\bibitem{MINOS}
P. Adamson {\it et al} [The MINOS+ Collaboration], arXiv:1710.06488v4[hep-ex].
\bibitem{gazal}
Gazal Sharma, Lal Singh and B. C. Chauhan, SESFC Conference Proc.(Abstract), 46 (2016).




 \end{thebibliography}
\end{document}